\documentclass[10pt,letterpaper]{article}
\usepackage[top=0.85in,left=2.75in,footskip=0.75in,marginparwidth=2in]{geometry}

\usepackage[utf8]{inputenc}

\usepackage{cite}

\usepackage[pdfencoding=auto]{hyperref}
\usepackage{bookmark}% faster updated bookmarks

\usepackage[right]{lineno}

\usepackage{microtype}
\DisableLigatures[f]{encoding = *, family = * }

\raggedright
\usepackage{changepage}

\usepackage[aboveskip=1pt,labelfont=bf,labelsep=period,singlelinecheck=off]{caption}

\makeatletter
\renewcommand{\@biblabel}[1]{\quad#1.}
\makeatother

\usepackage{lastpage,fancyhdr,graphicx}
\usepackage{epstopdf}
\fancyheadoffset[L]{2.25in}
\fancyfootoffset[L]{2.25in}

\usepackage{color}

\definecolor{Gray}{gray}{.25}

\usepackage{graphicx}

\usepackage{sidecap}

\usepackage{amsmath,amssymb}

\usepackage{wrapfig}
\usepackage[pscoord]{eso-pic}
\usepackage[fulladjust]{marginnote}
\reversemarginpar

\usepackage[nodayofweek]{datetime}
\newdateformat{mytoday}{\monthname[\THEMONTH] \twodigit{\THEDAY}, \THEYEAR}

\usepackage{multirow}
\usepackage{booktabs}
\usepackage{rotating}
\usepackage{tabulary}
\newcolumntype{K}[1]{>{\centering\arraybackslash}p{#1}}

\usepackage{pdflscape}
\usepackage{afterpage}
\usepackage{multirow}
\usepackage{booktabs}
\usepackage{url}
\usepackage{rotating}
\usepackage{amsmath}
\usepackage{amssymb}
\begin{document}
%\noindent Cite this preprint version of the manuscript as:

\ \\
%\noindent \textcolor{blue}{M. Mahmud, M.S. Kaiser, A. Hussain, S. Vassanelli. (2018). Applications of Deep Learning and Reinforcement Learning to Biological Data. \textit{IEEE Trans. Neural Netw. Learn. Syst.}, doi: 10.1109/TNNLS.2018.2790388.
\ \\
%\noindent \textcopyright\ IEEE holds the copyright of this work.
%}

\vspace*{0.35in}
% title goes here:
\begin{flushleft}
{\Large
%\textbf\newline{[Deep] [Reinforcement] Learning: An Overview and Applications to Biological Data}
\textbf\newline{Deep Learning in Mining Biological Data}
}
\newline
% authors go here:
\\
Mufti Mahmud\textsuperscript{1,*},
M. Shamim Kaiser\textsuperscript{2,*},
Amir Hussain\textsuperscript{3}
%,
% Author 5\textsuperscript{2},
% Author 6\textsuperscript{2},
% Author 7\textsuperscript{1,*}
\\
\bigskip
\textsuperscript{1}Dept. of Computing \& Technology, School of Science \& Technology, Nottingham Trent University, Nottingham, NG11 8NS, UK
\\
\textsuperscript{2} IIT, Jahangirnagar University, Savar, 1342 -  Dhaka, Bangladesh
\\
% \textsuperscript{3} Department of Biosensors and Biomedical Signals Processing, Silesian University of Technology, Gliwice, Poland\\
\textsuperscript{3} School of Computing, Edinburgh Napier University, Edinburgh, EH11 4BN, UK\\
\bigskip
\textsuperscript{*} Co-`first and corresponding' author. 
%\textsuperscript{*} 
Emails:  mufti.mahmud@ntu.ac.uk, muftimahmud@gmail.com (M. Mahmud); mskaiser@juniv.edu (M.S. Kaiser) \\
\bigskip
%Version: \mytoday \today; last saved: \currenttime~(GMT)

\end{flushleft}

\section*{Abstract}

Recent technological advancements in data acquisition tools allowed life scientists to acquire multimodal data from different biological application domains. Broadly categorized in three types (i.e., sequences, images, and signals), these data are huge in amount and complex in nature. Mining such enormous amount of data for pattern recognition is a big challenge and requires sophisticated data intensive machine learning techniques. Artificial neural network based learning systems are well known for their pattern recognition capabilities and lately their deep architectures - known as deep learning (DL) - have been successfully applied to solve many complex pattern recognition problems. Highlighting the role of DL in recognizing patterns in biological data, this article presents a comprehensive survey consisting of - applications of DL to biological sequences, images, and signals data; overview of open access sources of these data; description of open source DL tools applicable on these data; and comparison of these tools from qualitative and quantitative perspectives. At the end, it outlines some open research challenges in mining biological data and puts forward a number of possible future perspectives
% now start line numbers
%\linenumbers

% the * after section prevents numberin

\section*{Introduction}
\label{sec-intro}

Understanding pathologies, their early diagnosis and finding cures have driven the life sciences research in the last two centuries \cite{Coleman_biology_1977}. This accelerated the development of cutting edge tools and technologies that allow scientists to study holistically the biological systems as well as unprecedentedly dig down to the molecular details of the living organisms \cite{Magner_history_2002,Brenner_history_2012}. Increasing technological sophistication presented scientists with novel tools for DNA sequencing \cite{shendure_next-generation_2008}, gene expression \cite{metzker_sequencing_2010}, 
bioimaging \cite{Vadivambal_bioimaging_2016}, 
neuroimaging \cite{poldrack_progress_2015}, 
and brain-machine interfaces \cite{Lebedev-bmi-2017}. 

These innovative approaches to study the living organisms produce huge amount of data \cite{quackenbush_extracting_2007} and create a situation often referred as `Data Deluge' 
\cite{mattmann_computing_2013}. 
This biological big data can be characterized by their inherent characteristics of being \textit{hierarchical} (i.e., data coming from different levels of a biological system -- from molecules to cells to tissues to systems), \textit{heterogeneous} (i.e., data acquired by different acquisition methods -- from genetics to physiology to pathology to imaging), \textit{dynamic} (i.e., data changes as a function of time), and \textit{complex} (i.e., data describing nonlinear biological processes) \cite{li_big_2014}.
%The intrinsic complexity of the biological system combined with high-dimensionality, heterogeneity, and nonlinearity 
These intrinsic characteristics of the biological big data posed an enormous challenge to the data scientists to identify patterns and analyze them to infer meaningful conclusions from these data \cite{marx_biology_2013}. This triggered the development of rational, reliable, reusable, rigorous, and robust software tools  
%allowing the machines to intelligently detect patterns and analyze them in these data 
\cite{li_big_2014}
using machine learning (ML) based methods to facilitate recognition, classification, and prediction of patterns in the biological big data \cite{tarca_machine_2007}.

The conventional ML techniques can be broadly categorized in two large sets -- \textit{supervised} and \textit{unsupervised}. The methods pertaining to the \textit{supervised} learning paradigm classify objects in a pool using a set of known annotations, alternatively called attributes or features, i.e., learning from a few annotated data samples the remaining data are classified using those annotations. Instead, the techniques in the \textit{unsupervised} learning paradigm form groups (or clusters) among the objects in a pool by identifying their similarity, i.e., data annotations are first defined and then used for the data classification. Apart, there is a special category called \textit{reinforcement learning}, that allows a system to learn from the experiences it gains through interacting with its environment, and is out of the scope of this work.

%%%%%%%%%%%%%%%%% FIG1
%\begin{figure}[!htb]
\begin{figure*}[!htb]
\centering
\hspace*{-5cm} 
\includegraphics{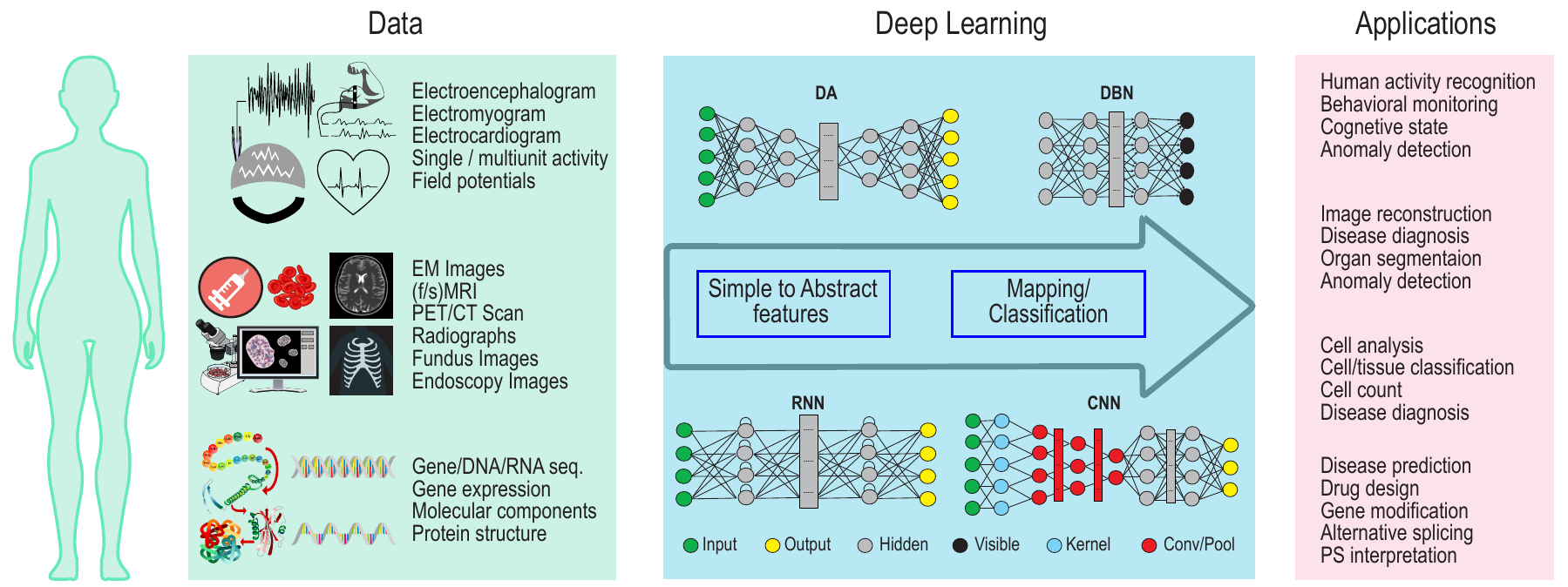}
\caption{Application of DL to biological data mining. The biological data coming from various sources (e.g., Omics, [Medical/Bio]-Imaging, [Brain/Body]-Machine Interfaces) are mined using DL with suitable architectures for specific applications.}
\label{fig_1}
\end{figure*}
%\end{figure}
%%%%%%%%%%%%%%%%%%%%%%%%%%

Some of the popular \textit{supervised} methods include: ANN and its variants, 
Support Vector Machines and other linear classifiers, 
Bayesian Statistics, k-Nearest Neighbors, Hidden Markov Model, and Decision Trees. On the other hand, a number of popular \textit{unsupervised} methods are: Autoencoders, Expectation-Maximization, Information Bottleneck, Self-Organizing Maps, Association Rules, Hierarchical Clustering, k-Means, Fuzzy Clustering, and Density-based Clustering.
Interested readers may refer to \cite{cheng_nn_review_1994,jain_ann_review_1996,kotsiantis_ml_review_2006} for brief introductory reviews on many of the techniques mentioned above.

The literature is in abundance with reports of successful application of the above mentioned popular ML methods and their respective variants to Biological data coming from various sources. For the sake of simplicity in this review, the Biological data sources have been categorized to a few broad application domains, e.g., \textit{Omics} (covers data from genetics and [gen/transcript/epigen/prote/metabol]omics \cite{horgan_omic_2011}), \textit{Bioimaging} (covers data from [sub-]cellular images acquired by diverse imaging techniques), \textit{Medical Imaging} (covers data from [medical/clinical/health] imaging mainly through diagnostic imaging techniques), 
and \textit{[Brain/Body]-Machine Interfaces or BMI} (covers mostly electrical signals generated by the Brain and the Muscles and acquired using appropriate sensors).
Each of these application domains (i.e., \textit{omics}
\cite{libbrecht_ml_2015}, 
\textit{bioimaging} 
\cite{kan_machine_2017},
\textit{BMI}
\cite{vidaurre_machine-learning-based_2010,mala_feature_2014,mahmud_processing_2016}, 
\textit{medical imaging} 
\cite{lemm_introduction_2011,erickson_machine_2017})
have witnessed major contributions from diverse ML methods (the ones mentioned above) and their variants.

In recent years Deep Learning (DL), Reinforcement Learning (RL), and deep RL methods are considered to reshape the future of ML (see the schematic diagram in Fig. \ref{fig_1} G)
\cite{mnih_human-level_2015}. 
Despite notable popularity and their applicability to diverse disciplines, there exists no comprehensive review in the literature focusing on Biological data. To fill this gap, this review provides-- a brief overview on DL, RL, and deep RL concepts; the state-of-the-art applications of these techniques to Biological data; and a comprehensive list of existing open source libraries and frameworks which can be utilized to harness the power of these techniques. Towards the end, some open issues are identified and some speculative future perspectives are outlined. Finally, working lists of available open access sources of datasets / databases from various application domains are supplied.

As for the organization of the rest of the article, section \ref{sec-overview} provides a conceptual overview to the DL technique and introduces the reader to the underlying theory; section \ref{sec-tools} presents the reader with brief descriptions of the popular open-source tools, software, and frameworks that implement DL techniques; section \ref{sec-perf-comp} provides a comparative study of the various tools' performances in implementing the defined DL architectures, 
section \ref{sec-issues-persp} provides the reader with some of the open issues and hints on the future perspectives; and finally, the article is concluded in section \ref{sec-conclusion}.

%------
\section{Overview of Deep Learning}
\label{sec-overview}
In DL the data representations are learned with increasing abstraction levels, i.e., at each level more abstract representations are learned by defining them in terms of less abstract representations at lower levels \cite{bengio_learning_2009}. Through this hierarchical learning process a system can learn complex representations directly from the raw data \cite{Goodfellow-et-al-2016}. 

Though many DL architectures have been proposed in the literature for various applications, the ones discussed below are most oftenly used in mining Biological data \cite{mahmud_DL_app_2017}.

\begin{figure}[!bth]
\includegraphics[scale=1]{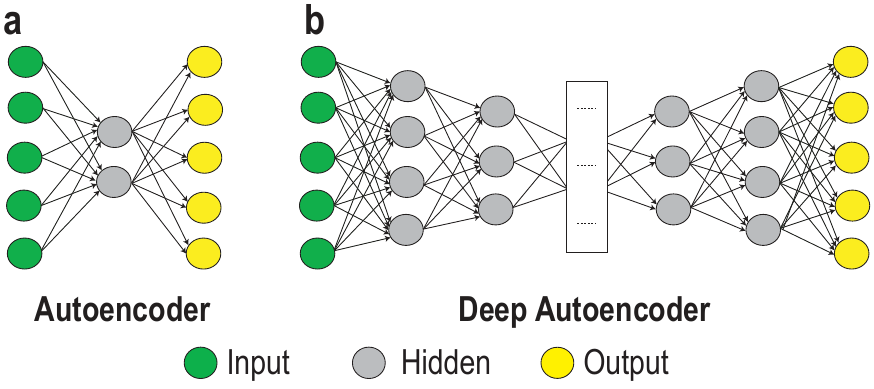}
\caption{Architectures of autoencoder (a) and deep autoencoder (b).}
\label{fig-da}
\end{figure}

\subsection{Autoencoder}
\label{subsec-da}
Autoencoder is a data driven unsupervised NN model mainly used for dimensionality reduction (see Fig. \ref{fig-da}a). It mainly projects high dimensional inputs to lesser dimensional outputs. In other words, an input (from $I$ input units) $x\in\mathbb{R}^{I}$ is mapped to a hidden unit (from $H$ hidden units) $a\in\mathbb{R}^{H}$ using a nonlinear activation function $f(\cdot)$ with $a=f(W^{(1)}x+b^{(1)})$, where $W^{(1)}\in\mathbb{R}^{H\times I}$ is the encoding weight matrix and $b^{(1)}\in\mathbb{R}^{H}$ is the bias vector. The projected $a$ is then reconstructed through remapping to an approximated value $y\in\mathbb{R}^{I}$  as $y=f(W^{(2)}a+b^{(2)})\approx x$, where $W^{(2)}\in\mathbb{R}^{I\times H}$ is the decoding weight matrix and $b^{(2)}\in\mathbb{R}^{I}$ is the bias vector. Usually Autoencoders use equal number of input and output units with lesser hidden units. However, to represent complex relationships among data, more hidden units with sparsity criteria have also been used. In both cases, the (non)linear transformations incorporated in the hidden units mainly perform the compression \cite{baldi_autoencoder_2012}. In the learning process, the goal of an autoencoder is to minimize the reconstruction error - for a given set of parameters - between $x$ and $y$. Thus, the objective function is given by:
\begin{equation}
E(X,Y)+\gamma\sum_j^{H}KL(\rho||\hat\rho_j),
\label{eq:ae-objf}
\end{equation}
where, $\gamma$ is the sparsity parameter, $KL(\rho||\hat\rho_j)=\rho \text{log}\frac{\rho}{\hat\rho_j}+(1-\rho)\text{log}\frac{1-\rho}{1-\hat\rho_j}$ is the relative entropy to measure how $j$th hidden unit's average activation ($\hat\rho_j$) diverges from target average activation ($\rho$), and $E(X,Y)=\frac{1}{2}\sum_{i=1}^{N}||x_i-y_i||_{2}^{2}$ is the reconstruction error for training set $\{X,Y\}=\{x_i,y_i\}_{i=1}^{N}$ with $N$ samples.

The Deep Autoencoder (DA) architecture, also known as `Stacked Autoencoder', (see Fig. \ref{fig-da}b) is obtained by stacking several Autoencoders where the activation values of one autoencoder's hidden unit become input to the next autoencoder, and backpropagation with gradient based algorithm is used to obtain the optimal weights. But this suffers from poor local minima problem which is overcome by pretraining the network with greedy layer-wise learning \cite{shen_dl_mia_2017}. 

Despite the pretraining stage and vanishing error problem \cite{bengio_vanishing-gradient_1994}, DA is a popular data compressing DL architecture with quite a few variants, e.g., 
Denoising Autoencoder \cite{vincent_denoising_autoencoders_2010}, 
Sparse Autoencoder \cite{ranzato_sparse_autoencoders_2006}, 
Variational Autoencoder \cite{kingma_variational_auto_encoding_2014}, and 
Contractive Autoencoder \cite{rifai_contractingauto_encoders_2011}.

\begin{figure}[!bthp]
\includegraphics[scale=1]{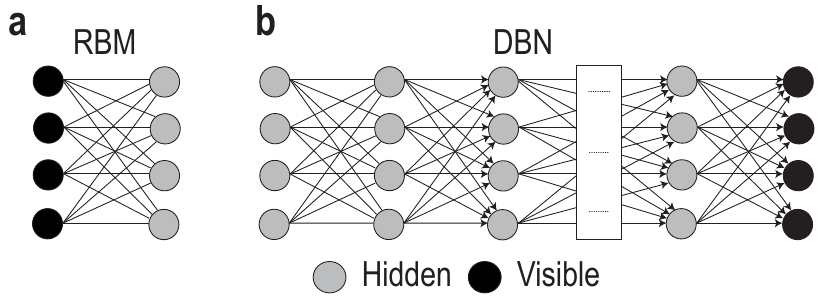}
\caption{Architectures of Restricted Boltzmann Machine (a) and Deep Belief Network (b).}
\label{fig-dbn}
\end{figure}

\subsection{Deep Belief Network}
\label{subsec-dbn}
Restricted Boltzmann Machine (RBM, Fig. \ref{fig-dbn}a), also considered as nonlinear feature detector, is an undirected probabilistic generative model capable of representing specific probability distributions \cite{Salakhutdinov_dbm_2009}. It contains one visible layer and one hidden layer with symmetric connections ($W\in\mathbb{R}^{V\times H}$) between them, with 
%$a\in\mathbb{R}^{V}$ and $b\in\mathbb{R}^{H}$ 
$a$ and $b$ as bias values for the visible and hidden layers, respectively. Generally, the visible layer contains $x\in\mathbb{R}^{V}$ units for the input observations, and the hidden layer contains $y\in\mathbb{R}^{H}$ units to model their relation with the observations. The symmetrical connections make RBMs usable as Autoencoders, and the joint probability of $(x,y)$ is given by \cite{zhou_dl_mia_book_2017}:
\begin{equation}
\label{eq:rbm-jntprb}
P(x,y;\Phi)=\frac{1}{Z(\Phi)}\textrm{exp}[-E(x,y;\Phi)],
\end{equation}
where, $\Phi=\{W,a,b\}$, $Z(\Phi)$ is a partition function derived from possible ($\forall x$, $\forall y$) pairs, and $E(x,y;\Phi)$ is the energy function which - for a generic case of binary visible and hidden units - is described as:
\begin{align}
\label{eq:rbm-energy}
E(x,y;\Phi) &= -y^\top W x - a^\top x - b^\top y \\ \nonumber
			&= -\sum_{i=1}^{V}\sum_{j=1}^{H}x_iW_{ij}y_j - \sum_{i=1}^{V}a_ix_i - \sum_{j=1}^{H}b_jy_j.
\end{align}
Here the conditional probability distributions of visible given hidden units and hidden given visible units are computed as - 
$P(x_i=1|y;\Phi)=\sigma(a_i+\sum_{j=1}^{H}W_{ij}y_j)$ and 
$P(y_j=1|x;\Phi)=\sigma(b_j+\sum_{i=1}^{V}W_{ij}x_i)$ 
respectively, with $\sigma(\cdot)$ as a logistic sigmoid function. Now, as the hidden units of RBM are unobservable, the objective function can be defined using the marginal distribution of the visible units only as:
\begin{equation}
\label{eq:rbm-objf}
P(x;\Phi)=\frac{1}{Z(\Phi)}\sum_{y}\textrm{exp}(-E(x,y;\Phi)).
\end{equation}
Training RBM parameters are done by maximizing the log-likelihood of the observations through a contrastive divergence algorithm. Gibbs sampling technique \cite{geman_gibbs_sampling_1984} is used to approximate the expected values of the distribution and calculate the gradient descent \cite{zhou_dl_mia_book_2017}.
%\cite{fischer_rbm_2012}. 

Stacking multiple RBMs as learning elements leads to a popular DL architectures known as Deep Belief Network (DBN, Fig. \ref{fig-dbn}b) where one RBM's latent layer is connected to the subsequent RBM's visible layer. Therefore, a DBN contains one visible layer $x$ and $L$ hidden layers $y^{\ell=1\ldots L}$. With downwards directed connections except the top two undirected layers, DBN is a hybrid model combining undirected graphical model and directed generative model \cite{hinton_dbn_2006}. 
The joint distribution of the visible units ($x$) and hidden layers ($y^{\ell=1\ldots L}$) is given by:
\begin{equation}
\label{eq:dbn-jntprb}
P(x,y^{\ell=1\ldots L})=\Big( \prod_{\ell=0}^{L-2}P(y^{\ell}|y^{\ell+1})\Big) P (y^{L-1},y^{L}).
\end{equation}
with $y^{0}=x$, and $P(y^{L-1},y^{L})$ denotes the joint distribution between layers $L-1$ and $L$.
Individual layers are pretrained in layerwise greedy fashion using unsupervised learning and perform generative fine tuning depending on the required outcome of the model \cite{ravi_dl_2017}. Nonetheless, the training process remains computationally expensive. 

\begin{figure}[!bthp]
\includegraphics[scale=1]{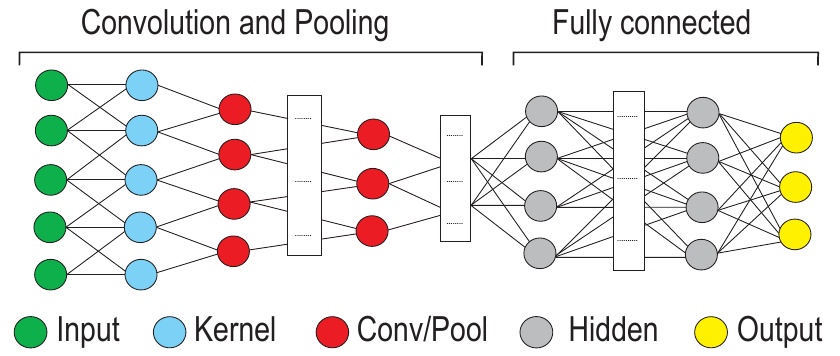}
\caption{Architecture of convolutional neural network.}
\label{fig-cnn}
\end{figure}

\subsection{Convolutional Neural Network}
\label{subsec-cnn}
CNN (Fig. \ref{fig-cnn}) is a multilayer NN model, comprised of convolutional layers (often interfused with subsampling layers) followed by fully connected layers, that mimics the locally sensitive, orientation selective neurons in the visual system \cite{lecun_cnn_1998}. CNN is designed to handle multidimensional locally correlated inputs, e.g. the 2D structure of an image or speech signal, and to avoid overfitting by sharing weights which also makes it easier to train with lesser parameters compared to a fully connected network with equal hidden units. These facilitated the wide usage of CNN in problems with large number of hidden units and training parameters. 

A convolutional layer recognizes local patterns in terms of features from the input feature maps through learnable filter kernels - $k_{ij}^{\ell}$
%$\ell$
. These convolution filters (CF) mainly represent connection weights between feature maps $i$ and $j$ belonging to the layers $\ell-1$ and $\ell$ respectively. The activations of a convolution layer's units ($A_j^{\ell}$) are computed by convolving activations of a vicinal subset of units from the preceding layer's feature maps ($A_i^{\ell-1}$) with the filter kernels ($k_{ij}^{\ell}$) as:
\begin{equation}
\label{cnn-actv}
A_j^{\ell}=f\Big( \sum_{i=1}^{N}A_i^{\ell-1}\circledast k_{ij}^{\ell}+b_j^{\ell}\Big),
\end{equation}
where $N$ is the total feature maps in $\ell-1$ layer, $\circledast$ is the convolution operator, $b_j^{\ell}$ is the bias at layer $\ell$, and $f(\cdot)$ is the nonlinear activation function \cite{bouvrie_notes_2006}.

A suitable pooling layer reduces the feature maps at every pooling step between the subsequent layers. These interspersed pooling layers, thus, reduces computational times and make CNN invariant to small spatial shifts. Also, because of the feature reduction at every applied step, only limited amount of features are eventually supplied to the fully connected network to classify.

When a convolutional layer $\ell$ is followed by a pooling layer $\ell+1$, a block of units in a feature map from layer $\ell$ are connected to a single unit of a feature map in layer $\ell+1$. 
The associated sensitivity map $\delta$ for layer $\ell$ is calculated as:
\begin{equation}
   \delta_j^{\ell} = f'(z_j^{\ell}) \bullet \text{up}\left(\delta_j^{\ell+1}\right),
\end{equation}
% \begin{align}
%    \delta_j^{\ell} = \text{upsample}\left((W_j^{\ell+1})^\top \delta_j^{\ell+1}\right) \bullet f'(z_j^{\ell}),
% \end{align}
where 
%$j$ indexes the filter number and 
$f'(\cdot)$ is the activation function's derivative evaluated using preactivations of convolutional layer $z_j^{\ell}$, and $\text{up}(\cdot)$ is the upsampling operation. 

When a current layer (pooling or covolutional) is followed by a convolutional layer, it is important to identify the correspondence in the feature maps between the two layers, i.e. the mapping between the current layer's patch and the next layer's unit in the feature maps. 
The gradients for the kernel wights are calculated using chain rule, and as the weights are being shared across multiple connections, they are given by:
\begin{equation}
  \frac{\partial E}{\partial k_{ij}^{\ell}} = \sum_{u,v}(\delta_j^{\ell})_{u,v}(P_i^{\ell-1})_{uv},
\end{equation}
where $(P_i^{\ell-1})_{uv}$ is the patch in the $i$th feature map ($A_j^{\ell-1}$) which is elementwise multiplied by the kernel ($k_{ij}^{\ell}$) during convolution to compute the element at $(u, v)$ in the output convolution feature map $A_j^{\ell}$ \cite{zhou_dl_mia_book_2017}.

Nonetheless, in case of very large datasets training even this kind of network can be daunting and can be solved using sparsely connected networks. Some of the popular CNN configurations include: AlexNet \cite{krizhevsky_alexnet_2012}, VGGNet \cite{simonyan_vgg_2014}, and GoogLeNet \cite{szegedy_googlenet_2015}.

\begin{figure}[!bthp]
\includegraphics[scale=1]{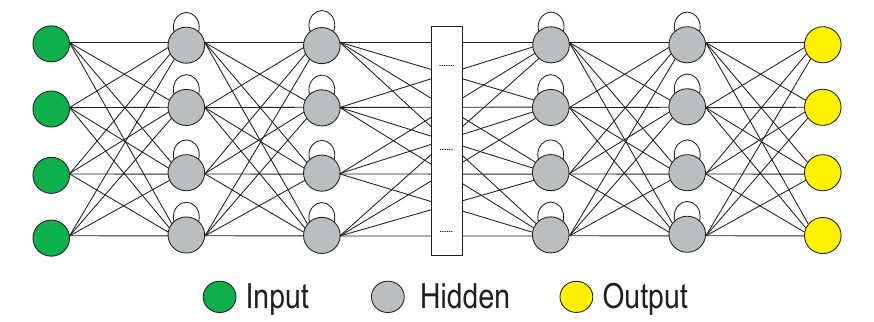}
\caption{Architecture of recurrent neural network.}
\label{fig-rnn}
\end{figure}

\subsection{Recurrent Neural Network}
\label{subsec-rnn}
RNN (Fig. \ref{fig-rnn}) is a NN model that detects sequences in streams of data. It computes the current state's output ($h_t$) for a given input ($x_t$) depending on the outputs of the previous states (captured by $h_{t-1}$) \cite{elman_finding_1990}:
\begin{equation}
h_t=f(Ux_t+Wh_{t-1}),
\end{equation}
where $f(\cdot)$ is a nonlinear function (e.g., tanh, ReLU \cite{zeiler_relu_2013}), and $U$ and $W$ are shared weight matrices. In other words, RNN learns a distribution over classes for a sequence of inputs (e.g., $x_1,x_2,\ldots,x_T$). As for the classification, generally a softmax, following few fully connected layers, is added for mapping the classes: 

$$(y|x_1,x_2,\ldots,x_T;\Phi= \text{softmax}(Vh_t),$$
%\begin{equation}
%	P⁡(y|x_1,x_2,\ldots,x_T;\Phi) = \text{softmax}(Vh_t),
%\end{equation}
where $V$ is the output weight matrix, and $\Phi$ is the set of parameters shared across different states.

Due to this `memory'-like property RNN gained popularity in many fields involving streaming data (e.g., text mining, time series, genomes, etc.). However, the backpropagating gradients- from the output through time- create learning problems similar to the conventional deep NN (e.g., vanishing and exploding gradients) \cite{lipton_critical_2015}.  In recent years, development of specialized memory units allowed expansion of classical RNN to useful variants including--  bidirectional RNN (BRNN) \cite{schuster_bidirectional_rnn_1997}, long short-term memory (LSTM) \cite{hochreiter_lstm_1997}, and gated recurrent units \cite{cho_grnn_2014}. Though RNN's primary application remains with sequential data, it is also increasingly applied to other data, e.g. images \cite{litjens_mia_2017}.
%have also been applied \cite{Lecun2015}.

\begin{table*}[!htbp]
  \centering
  \caption{Application of DL to Biological Data}
  \hskip-4.5cm
  \scriptsize
    \begin{tabular}{clcr}
    \toprule
    Source & Database/Dataset & DL Architecture & Application \\
%\cmidrule{3-6}          &       & \multicolumn{1}{l}{DA} & \multicolumn{1}{l}{DBN} & \multicolumn{1}{l}{CNN } & \multicolumn{1}{l}{RNN} &  \\
    \midrule
    \multirow{14}[14]{*}{\begin{sideways}Sequences\end{sideways}}
    & TCGA database      & DA \cite{danaee2016}      &Cancer detection \& gene identification
  \\
\cmidrule{2-4}          & Protein Data Bank (PDB)      & DA \cite{li_2016}
          & Protein structure reconstruction
 \\
\cmidrule{2-4}          & GWH \&  UCSC datasets     & DBN  \cite{Lee3045382}
     & Prediction of splice junctions
  \\
\cmidrule{2-4}          & SRBCT, Prostate Tumor, and MLL GE
     & DBN \cite{6944490}
             & Gene/MiRNA feature selection
  \\
\cmidrule{2-4}          &sbv IMPROVER       &  DBN \cite{Chen2015}
   &Human diseases \& drug development
  \\
\cmidrule{2-4}          &  iONMF dataset     & DBN \cite{pan_rbm_2017}
       & RNA-protein binding motifs
  \\
\cmidrule{2-4}          &ENCODE:       & CNN \cite{Denas2013DeepMO,Kelley2016}
 & Gene expression identification \\
\cmidrule{2-4}          &JASPAR database, \& ENCODE 
       &  CNN \cite{DBLPZengELG16}
      & Predicting DNA–protein binding 
 \\
\cmidrule{2-4}          &ENCODE DGF        &  CNN \cite{citeulike13721890}      & Predict noncoding-variant of Gene
 \\
\cmidrule{2-4}          & UCSC, CGHV Data, SPIDEX database          &  CNN  \cite{Huang069682}
       &genetic variants identification
  \\
\cmidrule{2-4}          & CullPDB, CB513, CASP datasets, CAMEO      &  CNN  \cite{DBLPWang0MX15} &2ps prediction \\

\cmidrule{2-4}          & DREAM 5       & CNN \cite{alipanahi_deepbind_2015} & DNA/RNA sequence prediction

  \\

\cmidrule{2-4}          & miRBoost      & RNN \cite{DBLPParkMCY16}
      & micro-RNA Prediction

  \\
\cmidrule{2-4}          & miRNA-mRNA pairing data repository     
& RNN (LSTM) \cite{DBLPLeeBPY16}
      &micro-RNA target prediction

  \\

    \midrule
    \multirow{14}[16]{*}{\begin{sideways}Images\end{sideways}} & MITOS dataset
      & CNN \cite{Ciresan2013}
       &Mitosis detection in breast cancer
  \\
\cmidrule{2-4}          &EM Segmentation Challenge   &  CNN  \cite{NIPS2012_4741}, RNN \cite{DBLPStollengaBLS15}      & Segment neuronal membranes, \\ & & & biomedical volumetric image 
 \\
\cmidrule{2-4}          &BRATS Dataset
      &  CNN \cite{DBLPHavaeiGLJ16}
 & Brain pathology segmentation
  \\
\cmidrule{2-4}          &ADNI MRI dataset
       &  CNN \cite{DBLPHosseiniAslGE16}, DBN \cite{Suk2014569,Li2014}
      & AD Diagnosis
  \\
\cmidrule{2-4}          &IBSR, LPBA40 \& OASIS dataset
       &  CNN \cite{Kleesiek2016460}
      & Skull stripping
  \\
\cmidrule{2-4}          &TBI dataset
      &  CNN \cite{kamnitsas_3dcnn_2017}
 &Brain lesion segmentation
  \\
\cmidrule{2-4}          &CT dataset  &  CNN \cite{Fritscher2016}
 &Fast segmentation of 3D medical images
  \\
\cmidrule{2-4}          & PACS Dataset
&  CNN \cite{DBLPChoLSCD15}
 &Medical image classification 
  \\
 \cmidrule{2-4}          &LIDC-IDRI dataset
 &  CNN  \cite{DBLPChoLSCD15}
  &Lung nodule malignancy classification
  \\
 \cmidrule{2-4}          & ADNI dataset
& DBN \cite{Suk2014569,Li2014}
      & AD/MCI diagnosis
  \\
 \cmidrule{2-4}          & ADHD-200      & DBN  \cite{7062868}
 & ADHD detection
  \\
 \cmidrule{2-4}          & MICCAI 2009 LV      & DBN \cite{ngo_dl_hrt_2017}
      & Heart LV segmentation
 \\
    \midrule
\multirow{7}[14]{*}{\begin{sideways}Signals\end{sideways}} &MAHNOB-HCI       & DA \cite{jirayucharoensak2014}  & Motion action decoding  \\
\cmidrule{2-4}          & BCI Competition IV            & DBN \cite{lu_rbm_mi_2016}, CNN \cite{yang2015,tabar_cnn_mi_eeg_2017, sakhavi_mi_2015}      &Motion action decoding

  \\
\cmidrule{2-4}          &DEAP dataset     & DBN \cite{li_dbn_as_2013,7033556}
   &Motion action decoding
  \\
\cmidrule{2-4}          &DEAP dataset      & CNN \cite{IAAI1715007}
   &Emotion classification
  \\
\cmidrule{2-4}          &Freiburg dataset
 & CNN \cite{Mirowski20091927}
    & Seizure prediction
  \\
\cmidrule{2-4}          & Ninapro database
   & DBN \cite{7023547}, RNN \cite{DBLPAtzoriCM16}
      & Motion action decoding
 \\
\cmidrule{2-4}          &MIT-BIH arrhythmia database 
  & DBN \cite{wu_ecg_dbn_2016,DBLPYanQWZ0W15}
      & Classification of ECG Arrhythmias
 \\
\cmidrule{2-4}          & MIT-BIH, INCART, \& SVDB
 & CNN \cite{DBLPAtzoriCM16}
       & Movement decoding
 \\
    \bottomrule
    \end{tabular}%
  \label{tab:appl}%
\end{table*}%

\section{Deep Learning and Biological Data}
\label{sec-apps}

% \textcolor{red}{
% This should contain the application of DL to biological data section from the TNNLS paper. We would need 3 information from each paper: 
% \begin{enumerate}
% \item data source used in a paper;
% \item architecture used in that paper;
% \item application of the analysis.
% \end{enumerate}
% In addition, we would need the `BibTex key' from the old TNNLS .bib file.
% \ \\
% A synthesized version of the application should be written by summarizing all the applied analysis and eventually creating a table \ref{tab:appl}}\\

Many studies have been reported in the literature which employ diverse DL architectures with related and varied parameter sets (see section \ref{sec-overview}) to analyze patterns in biological data. 
%The literature contains many studies which apply DL to analyze patterns in biological data. 
A summary of these studies which use open access data is reported in table \ref{tab:appl}.

\subsection{Sequences}
\label{subsec-app-seq}

Stacked Denoising DA was employed by Danaee et al. to extract features for cancer diagnosis and classification along with the identification of the related genes from GE data \cite{danaee2016}.
A template based DA learning model was proposed by Li et al. to reconstruct the protein structures \cite{li_2016}.
Lee et al. applied a DBN based unsupervised method to perform the auto-prediction of splicing junction at the level of DNA  \cite{Lee3045382}.
Combining DBN with active learning, Ibrahim et al. devised a method to select feature groups from genes or microRNAs (miRNAs)  based on expression profiles \cite{6944490}.
For translational research, bimodal DBNs were used by Chen et al. to predict responses of human cells using model organisms \cite{Chen2015}.
Pan et al. applied a hybrid CNN-DBN model on RNAs for the prediction of RNA binding protein (RBP) interaction sites and motifs \cite{pan_rbm_2017}, and Alipanahi et al. used CNN to predict sequence specificities of [D/R]BPs \cite{alipanahi_deepbind_2015}.
Denas and Taylor used CNN to preprocess ChIP-seq data and created gene transcription factor activity profiles \cite{Denas2013DeepMO}. 
CNN was used by Kelley et al. to predict DNA sequence accessibility \cite{Kelley2016}, by Zeng et al. to predict the binding between DNA and protein \cite{DBLPZengELG16}, by Zhou et al. \cite{citeulike13721890} and Huang et al.\cite{Huang069682} to find noncoding GV, and by Wang et al. to predict secondary protein structure \cite{DBLPWang0MX15}.
Park et al. used LSTM to predict miRNA precursor \cite{DBLPParkMCY16} and Lee et al. \cite{DBLPLeeBPY16} used it to predict miRNA precursors' targets.

\subsection{Images}
\label{subsec-app-images}

CNN was used by Ciresan et al. on histology images of the breast to find mitosis \cite{Ciresan2013} and Stollenga et al. used it to segment neuronal structures in Electron Microscope Images (EMI) \cite{NIPS2012_4741}.
Havaei et al. used CNN to segment brain tumor from MRI \cite{DBLPHavaeiGLJ16} and Hosseini et al. used it for the diagnosis of AD from MRI \cite{DBLPHosseiniAslGE16}.
DBM \cite{Suk2014569} and RBM \cite{Li2014} were used in detecting Alzheimer's Disease (AD) and Mild Cognitive Impairment (MCI) from MRI and PET scans. Again, CNN was used on MRI to detect neuroendocrine carcinoma \cite{Kleesiek2016460}. CNN's dual pathway version was used by Kamnitsas et al. to segment lesions related to tumors, traumatic injuries, and ischemic strokes \cite{kamnitsas_3dcnn_2017}. CNN was also used by Fritscher et al. for volume segmentation  \cite{Fritscher2016} and by Cho et al. to find anatomical structures (Lung nodule to classify malignancy) \cite{DBLPChoLSCD15} from CT scans.
DBN was applied on MRIs to detect Attention Deficit Hyperactivity Disorder \cite{7062868} and on cardiac MRIs to segment the heart's left ventricle \cite{ngo_dl_hrt_2017}.

% \noindent
% \includegraphics[scale=1]{Fig_Table_Format}

\subsection{Signals}
\label{subsec-app-signals}

Jirayucharoensak et al. used PCA to extract power spectral densities from each EEG channel, which were then corrected by covariate shift adaptation, finally stacked DA was used to detect emotion \cite{jirayucharoensak2014}.
DBN was applied to decode motor imagery (MoI) through classifying EEG signal frequency information \cite{lu_rbm_mi_2016}. For a similar purpose CNN was used covering large frequency ranges with augmented common spatial pattern features \cite{yang2015}. 
In a rather different approach using DA, features based on combined selective location, time, and frequency attributes were classified \cite{tabar_cnn_mi_eeg_2017}.
Li et al. used DBN to extract low dimensional latent features, and select critical channels to classify affective state using EEG signals \cite{li_dbn_as_2013}. Also, Jia et al. used an active learning to train DBN and generative RBMs for the classification \cite{7033556}.
Tripathi et al. utilized DNN and CNN based model for emotion classification using the DEAP dataset \cite{IAAI1715007}.
CNN was employed to predict seizures through synchronization patterns  classification from Freiburg dataset \cite{Mirowski20091927}.
DBN \cite{7023547} and CNN \cite{DBLPAtzoriCM16} were used to decode motion action from Ninapro database. The later approach was also used on MIT-BIH, INCART, \& SVDB repositories \cite{DBLPAtzoriCM16}.
Moreover, the ECG Arrhythmias were classified using DBN \cite{wu_ecg_dbn_2016,DBLPYanQWZ0W15} from the data supplied by MIT-BIH arrhythmia database.

% \noindent
% \includegraphics[scale=1]{Fig_Table_Format}

\section{Open Access Biological Data Sources}
\subsection{Omics}
\subsubsection{SGD}Saccharomyces Genome Database (SGD) provides complete biological information for the budding yeast \textit{Saccharomyces cerevisiae}. They also gives a open source tool for searching and analyzing these data, and thereby enable the discovery of functional relationships between sequence and gene products in fungi and higher organisms. The Study of Genome expression, transcriptome and computational biology are main function of the SGD.

\subsubsection{PubChem}The PubChem database contains millions of compound structures and descriptive datasets of chemical molecules and their activities against biological assays. Maintained by the National Center for Biotechnology Information of the United States National Institutes of Health, it can be freely accessed through a web user interface and downloaded via FTP. It also contains software services (such as plotting and clustering). It can be use for  [Gen/Prote]omics study and Drug design. 
\subsubsection{ENCODE}The Encyclopedia of DNA Elements (ENCODE) is a whole-genome database curated by the ENCODE Consortium which is composed primarily of scientists who were funded by US National Human Genome Research Institute. It contains genome datasets (including meta data) of human/mouse.

\subsubsection{MBD-UCI} Molecular Biology Databases (MBD) at the UCI
 contains three molecular biology databases: i) Protein Secondary Structure \cite{GE5}, which is a bench repository that classifies secondary structure of certain globular proteins; ii) Splice-Junction Gene Sequences \cite{GE6}, which contains primate splice-junction gene sequences (DNA) with associated imperfect domain theory; and iii) Promoter Gene Sequences \cite{GE7}, which contains E. Coli promoter gene sequences (DNA) with partial domain theory. Objectives- i) Sequencing and predicting the secondary structure of certain proteins; ii) Study primate splice-junction gene sequences (DNA) with associated imperfect domain theory; iii) Study E. Coli promoter gene sequences (DNA) with partial domain theory.
\subsubsection{INSDC} 
The International Nucleotide Sequence Database Collaboration \cite{GE11}, 
popularly known as INSDC, corroborates biological data from three major sources: i) DNA Databank of Japan \cite{GE11-1}, ii) European Nucleotide Archive \cite{GE11-2}, and iii) GenBank \cite{GE11-3}. These sources provide the spectrum of data raw reads, though alignments and assemblies to functional annotation, enriched with contextual information relating to samples and experimental configurations.

\subsubsection{NSD}
Nature Scientific data (NSD) includes omics data; taxonomy and species diversity; mathematical and modelling resources; cytometry; organism-focused resources and Health science data.This can be used for studying and modelling different aspect of Genomics

\subsubsection{SMPDB} The Small Molecule Pathway Database (SMPDB) includes 618 molecule pathways found in humans. 
This data are used for drug design, understanding gene / metabolite and protein complex concentration.
\subsubsection{TCGA database} 
The Cancer Genome Atlas (TCGA) contains more than two petabytes of genomic data of multi dimensional maps of prime genomic deviation in 33 categories of cancer. These data are generated by National Cancer Institute (NCI) and the National Human Genome Research Institute (NHGRI). This database is used to study genomic information for improving the prevention, diagnosis, and treatment of cancer.
 
\subsubsection{PDB} Protein Data Bank (PDB) contain more than 135 thousand data of proteins, nucleic acids, and complex assemblies. These can be used to understand all aspects of biomedicine and agriculture.  
\subsubsection{GEMS}
  Gene Expression Model Selector (GEMS) includes  microarray GeEx Data. Cancer Diagnosis and Biomarker Discovery are the two key objective of this dataset.
\subsubsection{CPD} Cancer Program Datasets (CPD) includes  Nearest Template Prediction (NTP), Parallel sequencing, Subclass Mapping (PSSM), DNA Microarray, gene sequence and different disease datasets. 

\subsubsection{Cancer GeEx}  Cancer gene expression (GeEx)   contains different cancer datasets which can be employed for designing tool/algorithm for 
disease detection
\subsubsection{iONMF dataset}
iONMF dataset contains Yeast RPR and RNA binding protein datasets. This datasets is used for analyzing multiple RNA binding proteins.

\subsubsection{JASPAR database}  JASPAR database is a database for transcription factor DNA binding profile.

\subsubsection{SysGenSim}
   SysGenSim includes bioinformatics tool, and Pula-Magdeburg single-gene knockout, StatSeq  and DREAM 5 benchmark datasets for studying Gene Sequence.
\subsubsection{miRBoost} The genomes of eukaryotes containing at least 100 miRNAs. This dateset is use for Studying post-transcriptional gene regulation (PTGeR)/miRNA-related pathology. 
\subsubsection{IGDD} The Indian Genetic Disease Database (IGDD) tracks of mutations in the normal genes for genetic 
diseases reported in India  Retrieve and study  genetic disorders is the main objective of this database.

\begin{table*}[!tbph]
\centering
\caption{Omics Databases/Datasets}
\label{Tableomics}
  \hskip-4.5cm
  \scriptsize
\begin{tabular}{m{0.8cm}m{2.4cm}m{7.7cm}m{5.2cm}}
\toprule 
\textbf{Ref.} & \textbf{Database/Dataset} & \textbf{Description} & \textbf{Target Application}  \\ \midrule
\cite{GE1} &\hspace*{0pt}\ignorespaces SGD& Provides biological data for budding yeast and analysis tool & Genome expression \&, transcriptome \\ \midrule

\cite{GE2} &PubChem  & Contains compound structures, molecular datasets and tool & [Gen/Prote]omics study \& Drug design  \\ \midrule
\cite{GE3} &ENCODE  & Encyclopedia of DNA Elements  & Genome study and their functions\\ \midrule

\cite{noauthor_uci_nodate} & MBD-UCI & Contains three molecular biology databases &  i) Sequencing and study gene sequences  \\ \midrule
\cite{noauthor_international_nodate} & INSDC &Includes nucleotide sequence data  & Study and analyze nucleotide sequence  \\ \midrule
\cite{GE12} & NSD & Includes omics and Health science data &Study different aspect of Genomics\\ \midrule
 \cite{GE13} & SMPD  & Includes 618 molecule pathways found in humans  & Drug design and understand GeEx\\ \midrule
  \cite{noauthor_cancer_nodate} &TCGA database   &Contains more than two petabytes of genomic data &Study genomic for cancer treatment.
 \\ \midrule
  \cite{noauthor_rcsb_nodate} &PDB   &Proteins, nucleic acids, and complex assemblies data &Understand all aspects of biomedicine  \\ \midrule
%  \cite{} &UCSC datasets   &  & \\ \midrule
  \cite{noauthor_gems:_nodate}&GEMS & Microarray Gene Expression Data&Cancer Diagnosis \& Biomarker Discovery\\ \midrule
  \cite{noauthor_cancerds_nodate} &CPD   & contains sequence and different disease datasets   & Disease detection \\ \midrule
  \cite{noauthor_bioinformatics_nodate} &Cancer GeEx   & It contains different cancer datasets  &  Disease detection \\ \midrule
 % \cite{ } &sbv IMPROVER   &  & \\ \midrule
 \cite{mstrazar_ionmf:_2017} &iONMF dataset   & It contains Yeast RPR and RNA binding protein datasets  & Analysis of RNA binding proteins\\ \midrule
\cite{noauthor_jaspar_nodate} &JASPAR database   &A database for transcription factor binding profile  & Study binding profile\\ \midrule
%\cite{} &CASP datasets   &  & \\ \midrule
\cite{noauthor_sysgensim_nodate} &SysGenSim   &Bioinformatics tools and gene sequence dataset & Gene Sequence study\\ \midrule
\cite{noauthor_mirboost_nodate} &miRBoost   & The genomes of eukaryotes containing at least 100 miRNAs   &Study PTGeR/miRNA- pathology. \\ \midrule
%\cite{} &miRNA-mRNA pairing data repository   &  & \\ \midrule
%\cite{} &CGHV Data, SPIDEX database   &  & \\ \midrule
\cite{GE14} &IGDD&Tracks of mutations in the normal genes & Retrieve and study  genetic disorders  \\ \bottomrule
\end{tabular}
\end{table*}

\subsection{Imaging}
\subsubsection{Image Science Database}
This database includes different Biological imaging database of acute lymphoblastic leukemia image, cell centeredimage iibrary, Euro-BioImaging BioSharing Collection, Hematology Images, Microscopic World Image Gallery, Molecular Expressions Photo Gallery, and Medical Imaging database such as Chest Radiograph, Mammography, MedPix Medical and Retinal Image etc.

\subsubsection{Bioimaging} It consists of radiogenomics,  genetic / chemical databases, and cell and tissue phenotypes databases and bioimage processing tools. The targeted applications: design algorithm for features extraction and anomaly detection.

\subsubsection{CellImageLibrary}
It presents cell image datasets and Cell Library app. The aim of this dataset is to study cell biology.

\subsubsection{BDTNP} Berkeley Drosophila Transcription Network Project (BDTNP) contains 3D Gene expression data, In-vivo DNA binding data as well as Chromatin Accessibility data (ChAcD).  Research on gene expression and detect anomaly are the key applications of this dataset. 

\subsubsection{EuroBioimaging} It provides biological and biomedical imaging data. The analysis of image data in bioimaging is the prime objective of this dataset. 
\subsubsection{CCDB}The Cell Centered Database (CCDB) provides API for high resolution 2/3/4D data from e-microscope and software tools to analyze the images.

\subsubsection{JCB Data Viewer} JCB Data Viewer facilitates viewing, analysis, and sharing of multi-D image data.for Analyzing cell biology. 

\subsubsection{MITOS dataset} MITOS dataset contains breast cancer histological images (haematoxylin and eosin stained slides). The Detection of mitosis and 
evaluation of nuclear atypia are key uses.   
\subsubsection{IBSR}The Internet Brain Segmentation Repository (IBSR) gives segmentation results of MRI data. Development of segmentation methods is the main application of this IBSR.. 
\subsubsection{LPBA40}  The LONI Probabilistic Brain Atlas (LPBA40) contains maps of brain anatomic regions of 40 human volunteers. 
Each map generates  a set of whole-head MRI whereas each MRI describes to identify 56 structures of brain, most of them lies in the cortex. The Study of skull-stripped MRI volumes, and classification of the native-space MRI, probabilistic maps are key uses of LPBA40. 
\subsubsection{ADHD 200} Attention Deficit Hyperactivity Disorder (ADHD) dataset includes 776 resting-state fMRI and anatomical datasets which are fused over the 8 independent imaging sites. The phenotypic information includes: age, sex, diagnostic status, measured ADHD symptom, intelligence quotient (IQ) and medication status.  Imaging-Based Diagnostic Classification is the main aim of ADHD 200 dataset.
\subsubsection{Open fMRI} It contains MRI \& EEG datasets to study brain regions and its functions. 
\subsubsection{OASIS}  The Open Access Series of Imaging Studies contains MRI datasets and open source data management platform (XNAT) to study and analyze Alzheimer’s Disease. 
\subsubsection{Neurosynth} Neurosynth includes fMRI literature (with some datasets) and synthesis platform to study Brain structure, functions and disease.
\subsubsection{UK data service} fMRI dataset contains fMRI datasets which can be useful for studying brain tumour surgical planning  
\subsubsection{ABIDE} The Autism Brain Imaging Data Exchange (ABIDE) includes autism brain imaging datasets for studying autism spectrum.  
\subsubsection{Open NI} Open Neuroimaging dataset contains imaging Modalities and brain diseases data which can be used to study decision support system for disease identification 
\subsubsection{NITRC} The Neuroimaging Informatics Tools and Resources Clearinghouse contains range of imaging data from MRI to PET, SPECT, CT, MEG/EEG and optical imaging for analyzing Functional and structural neuroimages.
\subsubsection{ADNI} Alzheimer's Disease Neuroimaging Initiative (ADNI) includes mild cognitive impairment (MCI), early AD and elderly control subjects diagnosis data. for detecting and tracking of Alzheimer’s disease
\subsubsection{Brain development} It provides neuroimaging data and toolkit software  to Identify  normal, healthy subjects.
\subsubsection{NeuroVault.org} This is a web-based repository (API) for collecting and sharing statistical maps of the human brain to Study human brain regions.
\subsubsection{TCIA}  The Cancer Imaging Archive (TCIA) contains CT, MRI, and nuclear medicine  (e.g. PET) images for Clinical diagnostic, biomarker and cross-disciplinary investigation. 

%%%%%%%%% BIORepository
\begin{table*}[!tbph]
\centering
\caption{Bioimaging Databases/Datasets}
\label{TableBioimaging }
  \hskip-4.5cm
  \scriptsize
\begin{tabular}{m{0.8cm}m{2.5cm}m{7.3cm}m{5.4cm}}
\hline 
\textbf{Ref.} & \textbf{Database/Dataset} & \textbf{Description} & \textbf{Target Application}  \\ \hline
%\cite{Bio1} &ImageJ& This contains The Cell Image Library & Study the cellular attributes \\ \hline
\cite{Bio2} &Image Science& Biological and medical imaging databases  & Cellular image analysis \& visualization \\ \hline
\cite{Bio3} &Bioimaging& Genetic/chemical and cell/tissue phenotypes databases& Feature extraction \& anomaly detection \\ \hline
\cite{Bio4} &CellImageLibraryCell &image datasets and Cell Library app. & Study cell biology \\ \hline
\cite{Bio5}  &BDTNP & 3D Gene expression , DNA binding data \& ChAcD & Gene expression and detect anomaly \\ \hline
\cite{Bio6}  &EuroBioimaging& biological and biomedical imaging data & Analyze bioimaging \\ \hline
\cite{Bio7}  &CCDB& API for high resolution 2/3/4D  EM data& Analyzes bioimage\\ \hline
\cite{Bio8}  &JCB Data Viewer& viewing, analysis, and sharing of multi-D image data. &  Analyse cell biology \\ \hline
\cite{noauthor_mitos-atypia-14_nodate} &MITOS dataset & Breast cancer histological images 
 & Evaluation of nuclear atypia   \\ \hline
%\cite{} &EM Segmentation Challenge &  &  \\ \hline
%\cite{} &BRATS Dataset &  &  \\ \hline
\cite{noauthor_nitrc:_nodate} &IBSR &Segmentation results of MRI data.  & Development of segmentation methods. 
 \\ \hline
\cite{shattuck_construction_2008} &LPBA40 &Maps of brain regions and a set of whole-head MRI. 
 &Study MRI and map brain region  \\ \hline
%\cite{} &TBI dataset &  &  \\ \hline
%\cite{} &CT dataset &  &  \\ \hline
%\cite{} &PACS Dataset &  &  \\ \hline
%  TCIA\cite{} &LIDC-IDRI dataset &  &  \\ \hline
\cite{noauthor_adhd200_nodate} &ADHD-200 & fMRI/anatomical datasets  fused over the 8 imaging sites &Imaging-Based Diagnostic Classification  \\ \hline
%\cite{} &MICCAI 2009 LV &  &  \\ \hline
\cite{NM2}  &OpenfMRI  & MRI \&EEG datasets & Study brain regions and functions  \\ \hline
\cite{NM3} &OASIS  & MRI datasets and XNAT data management platform&Alzheimer’s Disease Research   \\ \hline
\cite{NM4} &Neurosynth & fMRI datasets and synthesis platform& Brain structure, functions and disease \\ \hline
\cite{NM5} & UK data service &fMRI dataset  & Brain tumour surgical planning \\ \hline
\cite{NM6} & ABIDE  & Autism brain imaging datasets& Study autism spectrum  \\ \hline
\cite{NM7} &Open NI& Imaging Modalities and brain diseases data& study DSS for disease identification   \\ \hline
\cite{Nm8} & NITRC & MRI, PET, SPECT, CT, MEG/EEG and optical imaging &Functional/structural neuroimage analysis  \\ \hline
\cite{NM9} & ADNI & MCI, early AD and elderly control subjects' diagnosis data. & Early detection of Alzheimer’s disease  \\ \hline
\cite{NM10} &Brain development  & It provides neuroimaging data and toolkit software  & Identify  normal, healthy subjects \\ \hline
\cite{NM11} &NeuroVault.org  &  API for collecting and sharing statistical maps of brain & Study human brain regions  \\ \hline
\cite{NM12} & TCIA  & CT, MRI, and  PET images & Diagnoses and biomarker investigation  \\ \hline

\end{tabular}
\end{table*}

\subsection{[Brain/Body]-Machine Interfaces (BMI)}
\subsubsection{BCI Competition Dataset} The BCI Competition datasets include EEG datasets (such as Cortical negativity or positivity, feedback test trials, self-paced key typing, P300 speller paradigm, motor/mental imagery data, continuous EEG; EEG with eye movement), ECoG datasets (such as finger movement, motor/mental imagery data in EEG/ECoG) and MEG dataset(such as wrist movement). These datasets can be used for signal processing and classification methods for BMI. 

\subsubsection{DEAP} 
A Database for Emotion Analysis using Physiological Signals (DEEP) provides various datasets for analyzing the human affective states. 
The EEG and sEMG of 32 volunteers were generated while watching music videos to analyze the affective states
These volunteer also rated the video and The front face was also recorded for 22 volunteer with consent. 

\subsubsection{Ninapro} The NinaPro database includes  of the kinematic as well as the sEMG data of 27 subjects while these subjects were moving finger, hand and wrist. These data can be employed to study Biorobotics  

\subsubsection{UCI ML repository} This repository contains datasets of  using 2 lead ECG (m-HEALTH), ECG of heart-attacks patients, arrhythmia, 64 electrode EEG, 2 mental state (Relax), EMG of Lower Limb, sEMG  Brain decoding and anomaly detection are the focused application of this dataset.  

\subsubsection{Physionet} This sites contains neuroelectric and myoelectric databases (EEG, EHG, and  ECG databases), waveform databases, multi-parameter databases, CHB-MIT Scalp EEG Database, EOG datasets, EEG motor movement/imagery datasets, ERP based BCI recording. The MIT-BIH Supraventricular Arrhythmia Database, the Physionet Normal Sinus Rhythm Database (NSRDB), the Physionet Supraventricular Arrhythmia Database (SVDB) are also the part of Phyionet. Epileptic seizure onset detection and treatment, Modelling and development of the BMI instrumentation are some of the targeted applications of this database.
\subsubsection{BNCI Horizon 2020} This databse contains more than 25 datasets such as stimulated EEG datasets, ECoG-based BCI datasets, ERP-based BCI datasets, Mental arithmetic, motor imagery (extracted from EEG, EOG, fNIRS EMG) datasets,Neuroprosthetic control of an EEG/EOG datasets, speller datasets and so on. Modelling and designing of BMI devices are the key application of this database. 
\subsubsection{MAHNOB-HCI} MAHNOB-HCI datasets produces a ECG and EEG database for affect recognition and implicit tagging (stimulated by fragments of movies and pictures). 

\subsubsection{DECAF} DECAF is a multimodal dataset for decoding user physiological responses to affective multimedia content. It contains magnetoencephalogram (MEG), horizontal electrooculogram (hEOG), ECG, Trapezius muscle-EMG, near-infrared face video data to study Physiological and mental states. 
\subsubsection{Brain signals data} This datasets includes event-related potential (ERP), event-related synchronization (ERD), epileptic seizure studies, brain mapping (including fMRI data).  
\subsubsection{TELE ECG}TELE-ECG dataset includes 250 ECG records with annotated QRS and artifact masks. It also includes QRS and artifact detection algorithms to Study QRS and artifact detection from the ECG signal.

\subsubsection{LIMO EEG} This dataset includes Raw EEG data, and Group level covariate describing age of subjects and channel location describing all electrode. 
\subsubsection{ESSMN} This is a 128-channel EEG dataset which can be used to detect anomaly in the EEG signal.

\subsubsection{EEG}The EEG database contains invasive EEG recordings of 21 intractable focal epilepsy patients.

\subsubsection{Facial s-EMG} This is a 128-channel EEG data of single subject. This dataset can be used to study the Muscles potentials, 
%\subsubsection{BCI-ML FD} This dataset presents 

\begin{table*}[!tbhp]
\centering
\caption{BMI Open Access Data Sources}
\label{TableBMI}
  \hskip-4.5cm
  \scriptsize
\begin{tabular}{m{1cm}m{2.5cm}m{7.4cm}m{5cm}}
\hline 
\textbf{Ref.} & \textbf{Database/Dataset} & \textbf{Description} & \textbf{Target Application}  \\ \hline
\cite{BCI01} &BCI Competition & EEG,  ECoG and MEG dataset
   & Signal processing/ classification \\ \hline
\cite{BMI5} &DEAP &EMG/EEG data (while watching music/videos)   & Database for Emotion Analysis\\ \hline
\cite{noauthor_ninapro_nodate} &Ninapro database   &Kinematic as well as the sEMG data of 27 subjects& Study Biorobotics  \\ \hline
\cite{BMI1} &UCI ML repository & Various ECG, ECG, EMG, sEMG datasets   & Brain decoding and anomaly detection \\ \hline
\cite{BMI2} &Physionet  & Various recorded physiologic signals
  & seizure detection; Study BMI  \\ \hline
\cite{BMI3} &BNCIHorizon2020  & Various BMI signals datasets &Designing BMI devices \\ \hline
\cite{BMI4} &MAHNOB-HCI &   ECG/EEG database for affect recognition/implicit tagging & Affect Recognition study  \\ \hline
\cite{BMI6} &DECAF &MEG, hEOG, ECG, Trapezius muscle-EMG, face video data  & Study Physiological and mental states\\ \hline
\cite{BMI7} &Brain signals data & ERP, ERD, Epileptic seizure studies, Brain mapping & Seizure studies \& Brain mapping  \\ \hline
\cite{BMI8} &TELE ECG  & 250 ECG records with annotated QRS and artifact masks. & Study QRS and artifact detection.    \\ \hline
\cite{BMI9} &LIMO EEG & Raw EEG data of different age group & BMI study \\ \hline
\cite{BMI10} & ESSMN &A 128-channel EEG dataset  & Anomaly detection \\ \hline
\cite{BMI11} &EEG & Invasive EEG recordings of 21 intractable epilepsy patients& Study epilepsy \\ \hline
\cite{BMI12} &Facial s-EMG & This is a 128-channel EEG single subject  dataset & Muscles potential study \\ \hline
%\cite{BMI13} &BCI-ML FD  &  EEG data from 5 participants &Motor-related neural signal   \\ \hline
\end{tabular}
\end{table*}

\section{Open Source Deep Learning Tools}
\label{sec-tools}

Due to surging interest and concurrent multidisciplinary efforts towards DL in the recent years, several open source libraries, frameworks, and platforms are made available to the community. In the following sections, the popular open source tools, which aim to facilitate the technological developments for the community, are reviewed and summarized. This comprehensive list contains tools (also developed by individuals) which are well maintained with a reasonable amount of implemented algorithms. For the sake of brevity, the individual publication references of the tools are omitted and interested readers may consult them at their respective websites from the provided urls. 

Table \ref{tab:tools} summarizes the main features and differences of the various tools. To measure the impact and acceptability of a tool in the community, we provide GitHub based measures such as, numbers of Stars, Forks, and Contributors. These numbers are indicative of the popularity, maturity, and diffusion of a tool in the community. 

\begin{table*}[!htbp]
  \centering
  \caption{Summary of Open Source Deep Learning Tools}
    \hskip-4.5cm
  \scriptsize
    \begin{tabular}{|l|l|r|r|r|r|r|r|}
    %\toprule
    \hline
     \multicolumn{1}{|c|}{\textbf{Sl. No.}} & \multicolumn{1}{c|}{\textbf{Tool}} & \multicolumn{1}{c|}{\textbf{Platform}} & \multicolumn{1}{c|}{\textbf{Language(s)}} & \multicolumn{1}{c|}{\textbf{Stars*}} & \multicolumn{1}{c|}{\textbf{Forks*}} & \multicolumn{1}{c|}{\textbf{Contrib.*}} & \multicolumn{1}{c|}{\textbf{Supported DL Architecture}} \\
    %\midrule
    \hline
%    \multirow{17}[34]{*}{\begin{sideways}Deep Learning\end{sideways}} 
1 & Apache Singha$^1$ & L, M, W      & Py, C++, Ja      &  1117     & 258      & 30      & CNN, RNN, RBM, DBM \\
%\cmidrule{1-8}  
\cline{1-8}
2 & Caffe$^2$ & L, M, W, A      &  Py, C++, Ma     & 20858      &  12802     & 249      & CNN, RNN \\
%\cmidrule{1-8}  
\cline{1-8}
3 & Chainer$^3$ & L      &   Py    &  3063     & 814      & 140      &  DA, CNN, RNN \\
%\cmidrule{1-8}  
\cline{1-8}
4 & DeepLearning4j$^1$ & L, M, W      &   Ja    & 7465       &3717       &127       &  DA, CNN, RNN, RBM, LSTM\\
%\cmidrule{1-8}  
\cline{1-8}
5 & DyNet $^1$& L      &C++       & 1856      & 461       & 85      & CNN, RNN, LSTM \\
%\cmidrule{1-8}  
\cline{1-8}
6 & H$_2$O$^1$   & L, M, W       &Ja, Py, R       &2530	       &1027       &94       & CNN, RNN  \\
%\cmidrule{1-8}  
\cline{1-8}
7 & Keras$^3$ &  L, M, W     &Py       & 20822      &  7578     & 548      & CNN, RNN, DBN \\
%\cmidrule{1-8}  
\cline{1-8}
8 & Lasagne$^1$ & L, M      & Py      & 3266      & 906       & 62       & CNN, RNN, LSTM \\
%\cmidrule{1-8}  
\cline{1-8}
9 & MCT$^3$    &    W   &  C++     & 12817      & 3326      & 150      &CNN, DBN, RNN, LSTM  \\
%\cmidrule{1-8}  
\cline{1-8}
10 & MXNet$^1$ & L, M, W, A, I      &C++       & 11727      & 4325       & 437       & DA, CNN, RNN, LSTM \\
%\cmidrule{1-8}  
\cline{1-8}
11 & Neon$^1$  &  L, M     &   Py    &  3271      & 723     & 69      &  DA, CNN, RNN, LSTM \\
%\cmidrule{1-8}  
\cline{1-8}
12 & PyTorch$^2$ &  L, M     &     Py  & 8464      &   1762    &   330    &  CNN, RNN, LSTM \\
%\cmidrule{1-8}  
\cline{1-8}
13 & TensorFlow$^1$ &  L, M, W      &   Py, C++    & 74463       & 36781       & 1100      & CNN, RNN, RBM, LSTM   \\
%\cmidrule{1-8}  
\cline{1-8}
14 & TF.Learn$^3$ &   L, M    & Py, C++      &  6916     &  1513     &  107     &  CNN, BRNN, RNN, LSTM \\
%\cmidrule{1-8}  
\cline{1-8}
15 & Theano$^2$ &  L, M, W     &  Py     &  7171     &    2319   &   323    &  CNN, RNN, RBM, LSTM \\
%\cmidrule{1-8}  
\cline{1-8}
16 & Torch$^2$ & L, M, W, A, I       &	Lu, C, C++       &  7387     &  2174     &  134     & CNN, RNN, RBM, LSTM \\
%\cmidrule{1-8}  
\cline{1-8}
17 & Veles$^1$ &  L, M, W, A     &  Py     & 840       &  185     &  8     &  DA, CNN, RNN, RBM, LSTM\\
    %\bottomrule
    \hline
    \end{tabular}%
  \label{tab:tools}%
 
 \vspace{2ex}
 
 \raggedright *GitHub parameters (as of 25 Oct. 2017); $^1$Apache2 License; $^2$BSD License; $^3$MIT License; \\
% $^4$GNU-GPL License; $^5$Mozilla Public License; \\
 \textbf{Legends}: L--Linux/Unix; M--MacOSX; W--Windows; A--Android; I--iOS; CP--Cross-platform; Py--Python; Ja--Java; Lu--Lua; Ma--Matlab. 
%PG--Policy Gradient; QL--Q Learning; UCB--Upper Condition Bound; GBEM--Gradient-based Bellman Error Minimization; GGQ--Greedy GQ; SGQ--Softmax GQ; NAC--Natural Actor-Critic; LSTDQ($\lambda$)--Least Squares TD-Q($\lambda$); QVL--QV-Learning; TNRE --Truncated Natural; RE--Relative Entropy; TR--Trust Region; CE--Cross Entropy; RWR--Reward Weighted Regression; CMAES--Covariance Matrix Adaption Evolution Strategy. 
%\\\textbf{Note}: `--' denote that the toolbox is hosted elsewhere (e.g., google code), thus, GitHub parameters are unavailable.

\end{table*}%

%------------------ MM Version

% \subsection{Deep Learning}
% \label{subsec-dltools}

% Apache Singa
\subsection{Apache Singa}
\label{subsec-singa}
Known as Singa (\url{https://singa.incubator.apache.org/}), it is a distributed DL platform written in C++, Java, and Python.
%\cite{ooi_singa_2015}. 
It's flexible architecture allows synchronous, asynchronous, and hybrid training frameworks to run. It supports a wide range of DL architectures including CNN, RNN, RBM, and DBM.

% -- Caffe
\subsection{Caffe}
\label{subsec-caffe}
Caffe (\url{http://caffe.berkeleyvision.org/}) is scalable, written in C++ and provides bindings for Python as well as Matlab.
Dedicated for experiment, training, and deploying general purpose DL models, this framework allows switching between development and deployment platforms. Targeting computer vision applications, it is considered as the fastest implementation of the CNN. 

% -- Chainer
\subsection{Chainer}
\label{subsec-chainer}
Chainer (\url{http://chainer.org/}) is a DL framework provided as Python library. Besides  the availability of popular optimization techniques and NN related computations (e.g., convolution, loss, and activation functions), dynamic creation of graphs makes Chainer powerful.
It supports a wide range of DL architectures including CNN, RNN, and DA.

% -- Deeplearning4j (DL4J)
\subsection{DeepLearning4j}
\label{subsec-dl4j}
Deeplearning4j (DL4J, \url{https://deeplearning4j.org/}), written in Java with core libraries in C/C++, is a distributed framework for quick prototyping that targets mainly nonresearchers. Compatible with JVM supported languages (e.g., Scala/Clojure), it works on distributed processing frameworks (e.g., Hadoop and Spark). 
Through Keras (section \ref{subsec-keras}) as a Python API, it allows importing existing DL models from other frameworks.
It allows creation of NN architectures by combining available shallow NN architectures.

% -- Dynamic Neural Network Toolbox
\subsection{DyNet}
\label{subsec-dynet}
The DyNet library (\url{https://dynet.readthedocs.io/}), written in C++ with Python bindings, is the successor of `C++ neural network library'.
In DyNet, computational graphs are dynamically created for each training example, thus, it is computationally efficient and flexible. Targeting NLP applications, its specialty is in CNN, RNN, and LSTM.

% -- H20 - for R, Python, Java
\subsection{H\texorpdfstring{$_2$}{\texttwoinferior}O}
\label{subsec-h20}
H$_2$O (\url{www.h2o.ai}) is an ML software that includes DL and data analysis.
It provides a unified interface to other DL frameworks like, TensorFlow, MXNet, and Caffe. It also supports training of DL models (CNN and RNN) designed in R, Python, Java, and Scala. 

% -- Keras
\subsection{Keras}
\label{subsec-keras}
The Python based Keras (\url{https://keras.io/}) library is used on top of Theano or TensorFlow.
Its models can be imported to DL4J (section \ref{subsec-dl4j}). It was developed as a user friendly tool enabling fast experimentation, and easy and fast prototyping. Keras supports CNN, RNN, and DBN.

% -- Lasagne
\subsection{Lasagne}
\label{subsec-lasagne}
Lasagne (\url{http://lasagne.readthedocs.io}) DL library is built on top of Theano. It allows multiple input, output, and auxiliary classifiers. 
It supports user defined cost functions and provides many optimization functions.
Lasagne supports CNN, RNN, and LSTM.

% -- Microsoft Cognitive Toolkit
\subsection{Microsoft Cognitive Toolkit}
\label{subsec-mct}
Replacing CNTK, the Microsoft Cognitive Toolkit (MCT, \url{https://cntk.ai/}) is mainly coded in C++.
It provides implementations of various learning rules and supports different DL architectures including DNN, CNN, RNN, and LSTM.

% -- MXNet
\subsection{MXNet}
\label{subsec-mxnet}
MXNet (\url{https://mxnet.io/}) framework allows defining, training, and deploying deep NN (DA, CNN, RNN and LSTM) on a wide range of devices-- from cloud infrastructure to mobile or even embedded devices (e.g., Raspberry Pi). Written in C++, it is memory efficient and supports Go, JavaScript, Julia, Matlab, Perl, Python, R, and Scala.

% -- Neon
\subsection{Neon}
\label{subsec-neon}
Neon (\url{www.nervanasys.com/technology/neon/}) is a DL framework written in Python. It provides implementations of various learning rules, along with functions for optimization and activation. Its support for DL architecture includes CNN, RNN, LSTM, and DA.

% -- PyTorch
\subsection{PyTorch}
\label{subsec-pytorch}
PyTorch (\url{http://pytorch.org/}) provides Torch modules in Python. More than a wrapper, its deep integration allows exploiting the powerful features of Python. Inspired by Chainer, it allows dynamic network creation for variable workload, and supports CNN, RNN and LSTM.

% -- TensorFlow
\subsection{TensorFlow}
\label{subsec-tensorflow}
TensorFlow (\url{www.tensorflow.org}), written in C++ and Python, is developed by Google and supports very-large-scale deep NN.
Amended recently as `TensorFlow Fold', its capability to dynamically create graphs made the architecture flexible, 
allowing deployment to a wide range of devices (e.g., multi-CPU/GPU desktop, server, mobile devices, etc.) without code rewriting. Also contains a data visualization tool named TensorBoard and supports many DL architectures including CNN, RNN, LSTM, and RBMs.

%%%%%%%%%% Relative comparison figure is here for the appearance purpose only %%%%%%%%%%%%%%%%%%%%%%%%%%

\begin{figure*}[!bth]
\hspace*{-5cm} 
\includegraphics[scale=1]{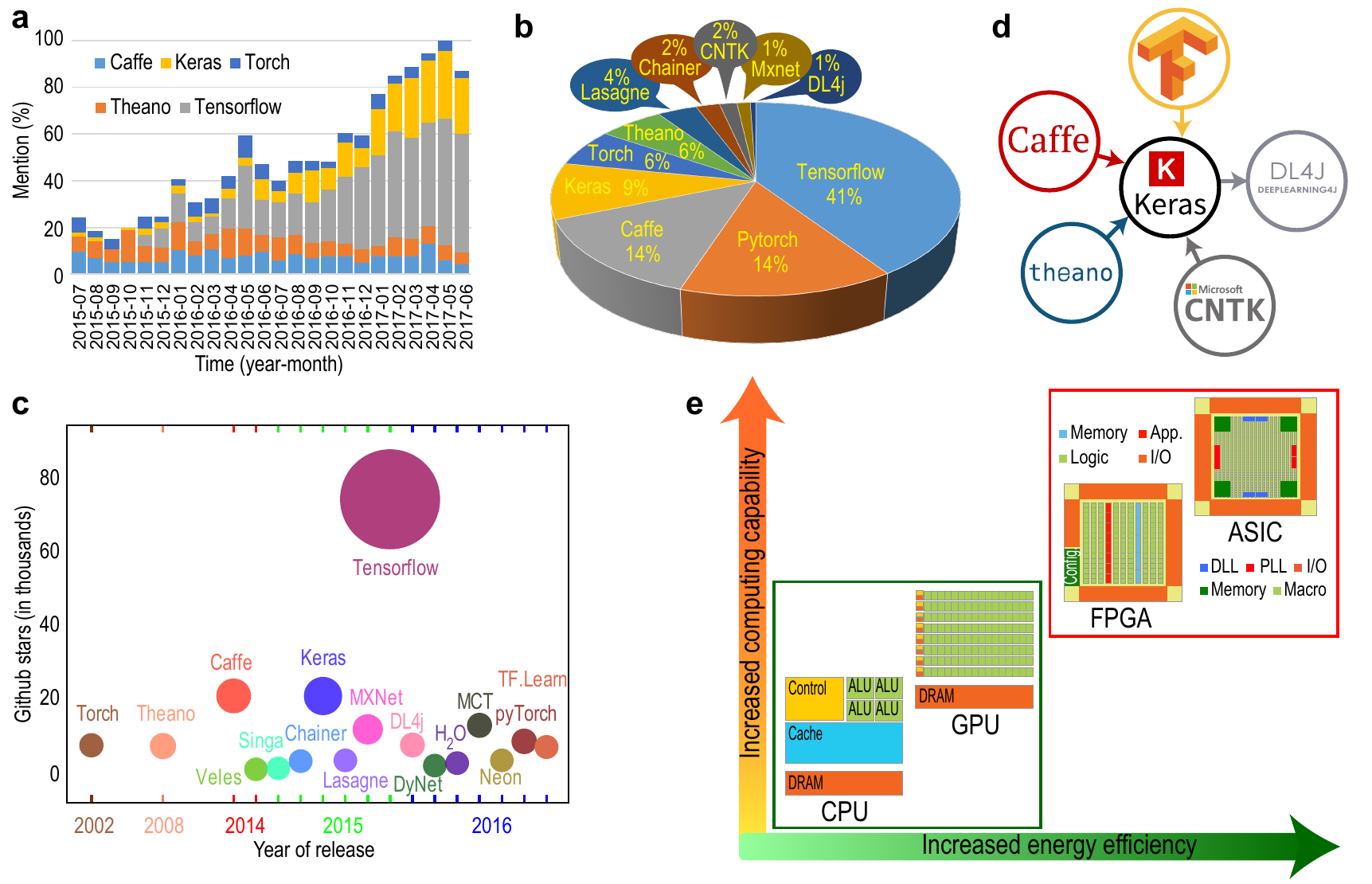}
\caption{Relative Comparison of DL tools. Popularity trend of individual DL tools as per mention in google search generated globally (a, data courtesy: google trend) and mention in arXiv's ML articles submitted during the month of March 2017 (b). 
The effect of community's participation on individual tools is shown by the bubble size, which is product of normalized number of GitHub forks and contributors (c).
As for the interoperability among the DL tools (d), Keras allows model importing from Caffe, MCT (CNTK), Theano, Tensorflow and lets DL4j to import.
Regarding hardware based scalability of the DL tools (e), most of the tools provide CPU and GPU support, whereas FPGA and ASIC can mainly execute pre-trained models.}
\label{fig-rel-comp}
\end{figure*}

% -- TFLearn
\subsection{TF.Learn}
\label{subsec-tflearn}
TF.Learn (\url{www.tflearn.org}) is a TensorFlow (section \ref{subsec-tensorflow}) based high level Python API.
It supports fast prototyping with modular NN layers and multiple optimizers, inputs, and outputs. Supported DL architectures include CNN, BRNN, and LSTM.

% -- Theano
\subsection{Theano}
\label{subsec-theano}
Theano (\url{www.deeplearning.net/software/theano/}) is a Python library that builds on core packages like NumPy and SymPy.
It defines, optimizes, and evaluates mathematical expressions with tensors, and served as foundation for many DL libraries. 

% -- Torch
\subsection{Torch}
\label{subsec-torch}
Started in 2000, Torch (\url{http://torch.ch/}), a ML library and scientific computing framework, has evolved as a powerful DL library.
Core functions are implemented in C and the rest via LuaJIT scripting language made Torch super fast. Software giants like Facebook and Google use Torch extensively. Recently Facebook's DL modules (fbcunn) focusing on CNN have been open-sourced as a plug-in to Torch.

% -- Veles
\subsection{Veles}
\label{subsec-veles}
Veles (\url{https://velesnet.ml/}) is a Python based distributed platform for rapid DL application development. It provides machine learning and data processing services and supports IPython notebooks.
Developed by Samsung, one of its advantages is that, it supports OpenCL for cross-platform parallel programming, and allows execution across heterogenous platforms (e.g., servers, PC, mobile, and embedded devices). The supported DL architectures include-- DA, CNN, RNN, LSTM, and RBM.

\section{Relative Comparison of DL Tools}
\label{sec-rel-comp}

To perform relative comparison among the available open-source DL tools, we selected four assessing measures for the tools which are detailed below: trend in their usage, community participation in their development, interoperability among themselves, and their scalability (see Fig. \ref{fig-rel-comp}).

% \begin{figure}[!bthp]
% \includegraphics[scale=1]{Fig_popularity}
% \caption{Popularity trend of individual DL tools as per - (a) mention in arXiv's ML articles submitted during the month of March 2017; (b) mention in google search generated globally as per google trend data.}
% \label{fig-trend}
% \end{figure}

\subsection{Trend}
\label{subsec-rel-trend}

To assess the popularity and trend of the various DL tools among the DL consumers, we looked into two different sources to assess the utilization of the tools. Firstly, we extracted globally generated search data from Google Trends\footnote{https://trends.google.com/} for two years (July 2015 to June 2017) related to search terms consisting of \big \langle[tool name] + Deep Learning\big \rangle. The data showed a progressive increase of search about Tensorflow since it's release followed by Keras (see Fig. \ref{fig-rel-comp}a).
Secondly, mining the content of around 2,000 papers submitted to arXiv's cs.[CV$|$CL$|$LG$|$AI$|$NE], and stat.ML categories, during the month of March 2017, for the presence of the tool names \cite{karpathy_peek_2017}. As seen in Fig. \ref{fig-rel-comp}b
%\ref{fig-trend}a, 
which shows an weighted percentage of each individual tool's mention in the papers, the top 6 tools were identified as: Tensorflow, Pytorch, Caffee, Keras, Torch, and Theano. 
%\ref{fig-trend}b).

%\cite{piatetsky_emerging_2017}

% \begin{figure}[!bthp]
% \includegraphics[scale=1]{Fig_Community}
% \caption{Community's support for the DL tools. The effect of community's participation on individual tools is shown by the bubble size, which is product of normalized number of github forks and contributors.}
% \label{fig-community}
% \end{figure}

\subsection{Community}
\label{subsubsec-rel-commu}

The community based development score for each tool discussed in Section \ref{sec-tools} was calculated from repository popularity parameters of GitHub (https://github.com/) (i.e., star, fork, and contributors). The bubble plot shown in Fig. \ref{fig-rel-comp}c
%\ref{fig-community} 
depicts community involvement in the development of the tools indicating the year of initial stable release. Each bubble size in the figure, pertaining to a tool, represents the normalized combined effect of fork and contributors of that tool. It is clearly seen that a very large part of the community effort is concentrated on Tensorflow, followed by Keras and Caffe.

% \begin{figure}[!bthp]
% \centering
% \includegraphics[scale=1]{Fig_Interoperability}
% \caption{Interoperability among the DL tools. Keras allows model importing from Caffe, MCT (CNTK), Theano, Tensorflow and lets DL4j to import.}
% \label{fig-interop}
% \end{figure}

\subsection{Interoperability}
\label{subsubsec-rel-iop}
In today's cross-platform development environments, an important measure to judge a tool's flexibility is it-s interoperability with other tools. In this respect, Keras is the most flexible one whose high-level neural networks are capable of running on top of either Tensor or Theano.  Alternatively, DL4j-model imports neural network models originally configured and trained using Keras that provides abstraction layers on top of TensorFlow, Theano, Caffe, and CNTK backends (see Fig. \ref{fig-rel-comp}d).
%\ref{fig-interop}).

% \begin{figure}[!b]
% \centering
% \includegraphics[scale=1]{Fig_Hardware}
% \caption{Hardware based scalability of the DL tools. Most of the tools provide CPU and GPU support. FPGA and ASIC can mainly execute pre-trained models.}
% \label{fig-hardware}
% \end{figure}

\subsection{Scalability}
\label{subsubsec-rel-cgpu}

Hardware based scalability is an important feature of the individual tools (see Fig. \ref{fig-rel-comp}e). Today's hardware for computing devices are dominated by graphics processing units (GPUs) and central processing units (CPUs). But considering increased computing capacity and energy efficiency, the coming years are expected to witness expanded role for other chipset types including application specific integrated circuits (ASICs), and field programmable gate arrays (FPGAs). So far DL has been predominantly used through software. Requirement for hardware acceleration, energy efficiency, and higher performance allowed development of chipset based DL systems.

\begin{figure*}[!bth]
\centering
\hspace*{-5cm} 
\includegraphics[scale=1]{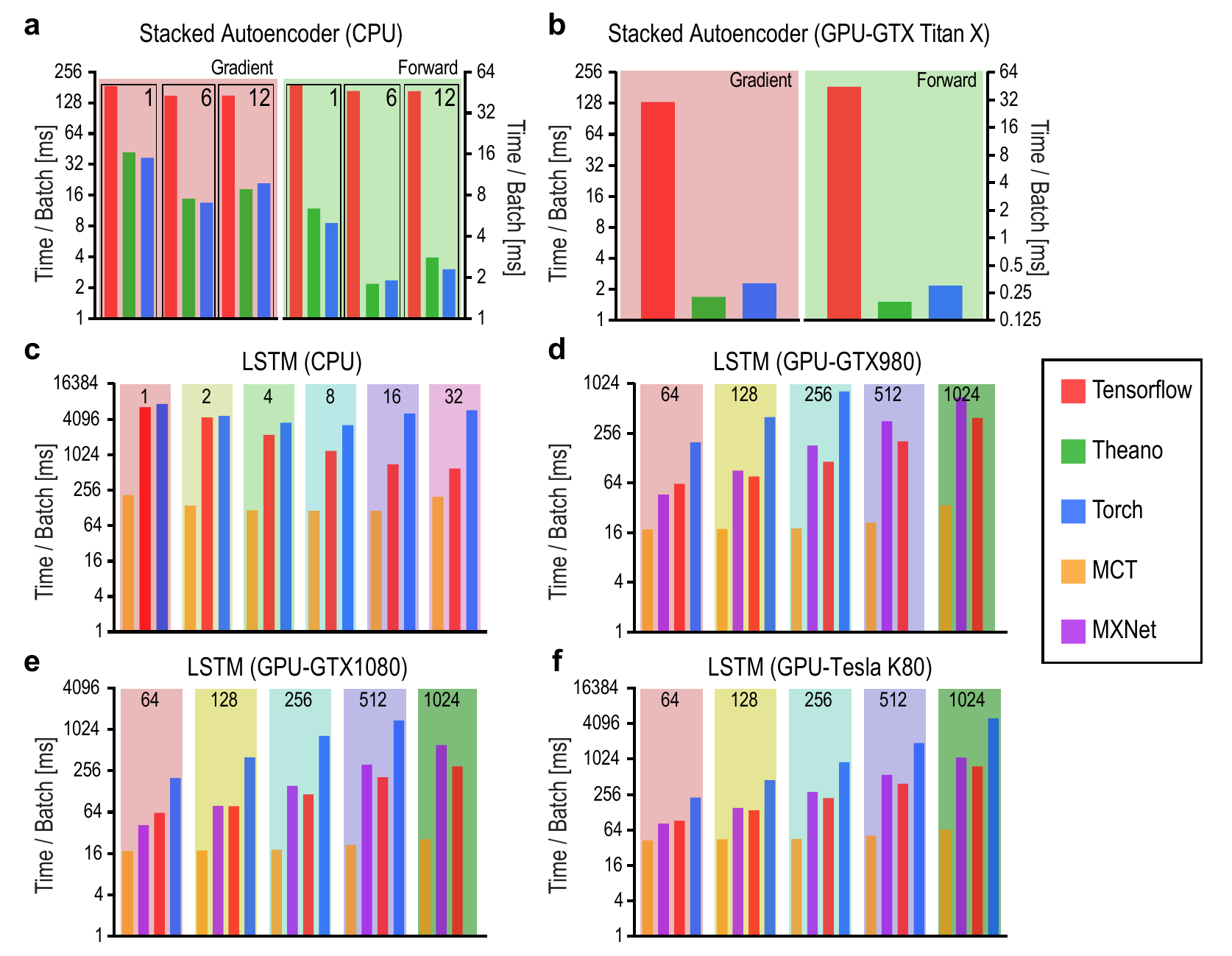}
\caption{Benchmarking Stacked Autoencoder or DA (a, b) and LSTM (c-f) in CPU and GPU platforms. The numbers in (a, c) denote the number of CPU threads employed in the benchmarking process, and in (d-f) denote the batch size. In case of DA the batch size was 64.} 
\label{fig-perf-DA-LSTM}
\end{figure*}

\section{Performance of Tools and Benchmark}
\label{sec-perf-comp}
%\textcolor{blue}{
The power of DL methods lie in their capability to recognize patterns for which they are trained. Despite the availability of several accelerating hardware (e.g., multicore [C/G]PUs), this training phase is very time consuming, cumbersome, and computationally challenging. Moreover, as each tool provides implementations of several DL architectures and often emphasizing separate components of them on different hardware platforms, selecting an appropriate tool suitable for an application is getting increasingly difficult. Besides, different DL tools have different targets, e.g., Caffe aims applications, whereas, Torch and Theano are more for DL research. To facilitate the scientists in picking the right tool for their application, a handful of scientists benchmarked the performances of the popular tools concerning their training times \cite{bahrampour_dl_fws_2016,shi_benchmarking_dl_2016}. Moreover, to the best of our knowledge, there exist two main efforts that provide the benchmarking details of the various DL tools and frameworks publicly \cite{deepmark_benchmark_2017,hk_benchmark_2017}. Summarizing those seminal works, below we provide the time required to complete the training process as a performance measure of four different DL architectures (e.g., FCN, CNN, RNN, and DA) among the popular tools (e.g., Caffe, CNTK, MXNET, Theano, Tensorflow, and Torch) on multicore [C/G]PU platforms.
%}

\begin{table}[!hb]
\scriptsize
  \centering
  \caption{Hardware configuration of the evaluating setup}
      \begin{tabular}{|l|l|c|}
    \hline
    \textbf{ESN} &\textbf{Processor} & \textbf{Memory}  \\  \hline
 \multirow{ 2}{*}{1} & \textbf{CPU:} E5-1650$^1$ @ 3.50 GHz &32 GB \\  \cline{2-3}
 %\cmidrule{2-4}
 & \multicolumn{2}{l|}{\textbf{GPU:} Nvidia GeForce GTX Titan X$^2$} \\ \hline
 \multirow{ 4}{*}{2} & \textbf{CPU:} E5-2630$^3$ @ 2.20 GHz &128 GB \\ \cline{2-3}
  & \multicolumn{2}{l|}{\textbf{GPU:} Nvidia GeForce GTX 980$^4$} \\ \cline{2-3}
   & \multicolumn{2}{l|}{\textbf{GPU:} Nvidia GeForce GTX 1080$^5$} \\ \cline{2-3}
 & \multicolumn{2}{l|}{\textbf{GPU:} Tesla K80 accelerator with GK210 GPUs$^6$} \\ \hline
  \multirow{ 4}{*}{3} & \textbf{CPU:} E5-2690$^3$ @ 2.60 GHz &256 GB \\ \cline{2-3}
 & \multicolumn{2}{l|}{\textbf{GPU:} Tesla P100 accelerator$^7$} \\ \cline{2-3}
  & \multicolumn{2}{l|}{\textbf{GPU:} Tesla M40 accelerator$^8$} \\ \cline{2-3}
   & \multicolumn{2}{l|}{\textbf{GPU:} Tesla K80 accelerator with GK210 GPUs$^6$} \\ \hline
 \end{tabular}%
  \label{tab:hardware}%
 \vspace{1ex}
 
 \raggedright 
 \textbf{Legends}: ESN: Experimental Setup Numbers; $^1$: Intel Xeon CPU v2; \ \ $^2$: 3072 cores, 1000 MHz base clock, 12 GB memory; $^3$: Intel Xeon CPU v4; $^4$: 2048 cores, 1126 MHz base clock, 4 GB memory; $^5$: 2560 cores, 1607 MHz base clock, 8 GB memory; $^6$: Tesla K80 accelerator has two Tesla GK210 GPUs with 2496 cores, 560 MHz base clock, 12 GB memory; $^7$: 3584 cores, 1189 MHz base clock, 16 GB memory; $^8$: 3072 cores, 948 MHz base clock, 12 GB memory.
\end{table}%

Table \ref{tab:hardware} lists the experimental setups used in benchmarking the specified tools. Mainly three different setups, each with Intel Xeon E5 CPU, were utilized during the process. Though the CPU were similar, the GPU hardware were different: GeForce GTX Titan X, GTX 980, GTX 1080, Tesla K80, M40, and P100.

\begin{figure*}[!bth]
\centering
\hspace*{-5cm} 
\includegraphics[scale=1]{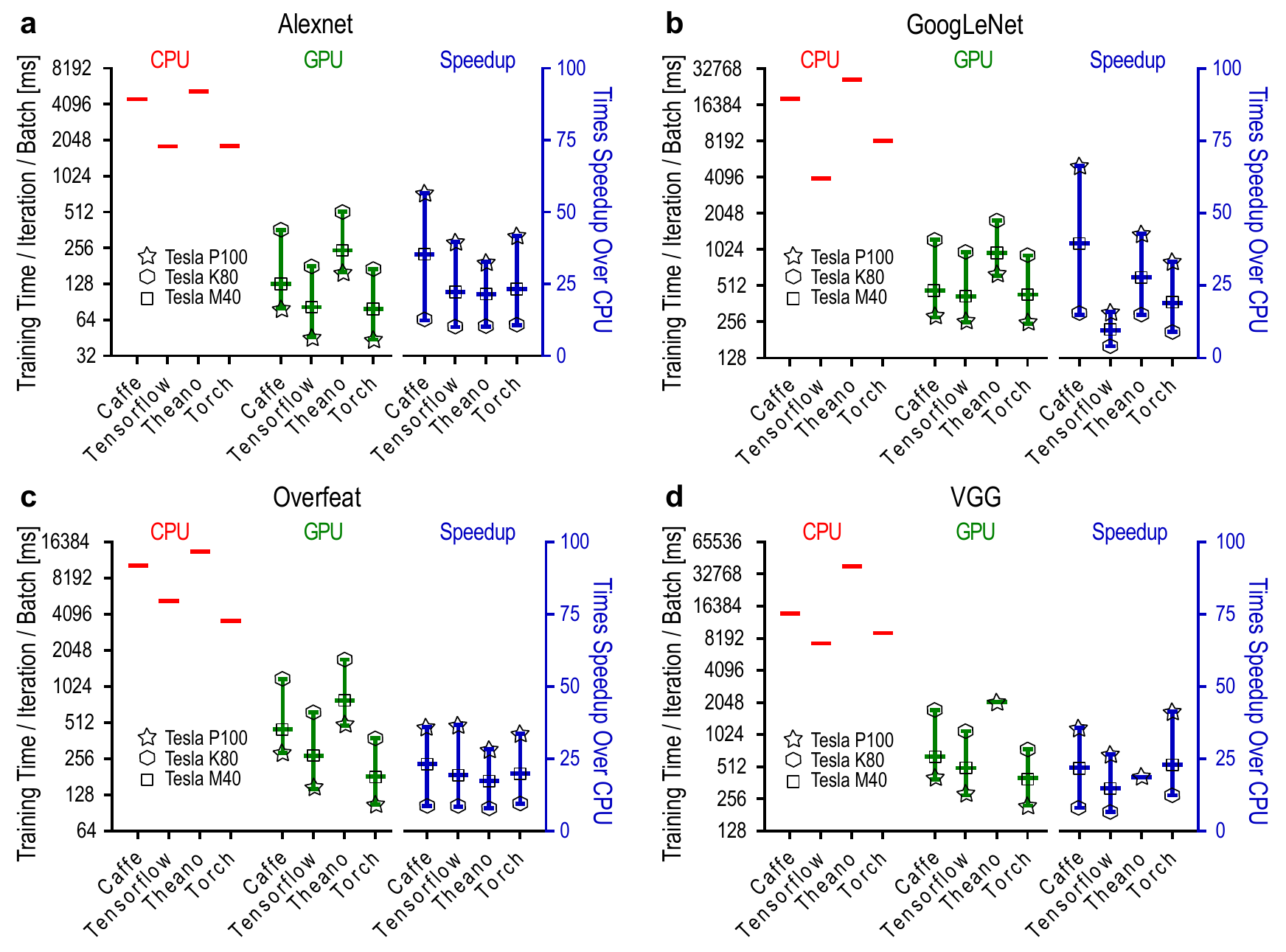}
\caption{The speedup of CNN training in different DL tools across various GPUs in comparison to CPU. The reported values were calculated for a batch size of 128, except for VGG for which the batch size was 64.} 
\label{fig-perf-CNNSpeedup}
\end{figure*}

Stacked autoencoders or DA were benchmarked using the experimental setup number 1 in Table \ref{tab:hardware}. To estimate the performance of the various tools on implementing DA, three autoencoders (number of hidden layers: 400, 200, and 100, respectively) were stacked with tied weights and sigmoid activation functions. A two step network training was performed on the MNIST dataset \cite{lecun_mnist_1998}. As reported in Fig. \ref{fig-perf-DA-LSTM} (a, b), the performances of various DL tools are evaluated using forward runtime and training time. The forward runtime refers to the required time for evaluating the information flow through the full network to produce the intended output for an input batch, dataset, and network. In contrast, the gradient computation time measures the time that required to train DL tools. The results suggest that, regardless of the number of CPU threads used or GPU, Theano and Torch outperforms Tensorflow both in gradient and forward times (see Fig. \ref{fig-perf-DA-LSTM} a, b).

Experimental setup number 2 (see Table \ref{tab:hardware}) was used in benchmarking RNN. The adapted LSTM network \cite{zaremba_rnn_2014} was designed with 10000 input and output units with two layers and $\sim$13 millions parameters. As the performance of RNN depends on the input length, an input length of 32 was used for the experiment. As the results indicate (see Fig. \ref{fig-perf-DA-LSTM} c-f), MCT outperforms other tools on both CPU and all three GPU platforms. On CPUs, Tensorflow performs little better than Torch (see Fig. \ref{fig-perf-DA-LSTM} c). On GPUs, Torch is the slowest with Tensorflow and MXNet performing similarly (see Fig. \ref{fig-perf-DA-LSTM} d-f).

% \begin{table}[htbp]
%   \centering
%   \caption{Version and GitHub Commit IDs of Evaluated software}
%       \begin{tabular}{|l|c|c|}
%     \hline
%     \textbf{Software} & \textbf{Version} & \textbf{GitHub Commit ID}  \\     \hline
%  Caffe& &8c8e832\\     \hline   
% % Neon&1.0.0.rc1  &a6766ff\\     \hline
% MCT &1.72 &f686879 \\ \hline
% MXNet &0.7 &34b2798 \\ \hline
% TensorFlow &0.11 &47dd089\\     \hline   
% Theano &0.7.0.dev &662ea98\\     \hline  
% Torch &7.0 &8c8e832\\     \hline   
% % openCV& 3.0&3684706\\     \hline   
% % OpenBLAS&0.2.14 &3684706\\     \hline   
 
%  \end{tabular}%
%   \label{tab:software}%
% \end{table}%

% \subsection{Speed}
% \label{subsec-perf-speed}
% \subsubsection{CNN}

% \begin{figure}[!bthp]
% \centering
% \includegraphics[scale=0.6]{CNN.pdf}
% \caption{The processing times of the (a) LeNet (b)AlexNet for both Gradient and forwarding pass.The batch size is 64} 
% \label{fig-perf-CNN}
% \end{figure}

%\subsubsection{AE}

% \begin{figure}[!bthp]
% \includegraphics[scale=0.45]{AE.pdf}
% \caption{The processing times of the stacked AE (Batch size=64) for both Gradient (including unsupervised pre-training and supervised fine-tuning ) and forwarding steps. The dimension of AE stacks are ( 800, 1000, and 2000) and (400,200,100). The tied weight can bot be used in DL tools such as Caffe and Neon}
% \label{fig-perf-AE}
% \end{figure}

%\subsubsection{LSTM}

% \begin{figure}[!bthp]
% \centering
% \includegraphics[scale=0.45]{STM.pdf}
% \caption{The processing times of the LSTM (Batch size=16) for both Gradient and forwarding pass.} 
% \label{fig-perf-LSTM}
% \end{figure}

Still a large portion of the pattern analysis is done using CNN, therefore, we further focused on CNN and investigated how the leading tools performed and scaled in training different CNN networks in different GPU platforms. Time speedup of GPU over CPU is considered as a metric for this purpose. The individual values are calculated using the benchmark scripts of DeepMark \cite{deepmark_benchmark_2017} on experimental setup number 3 (see Table \ref{tab:hardware}) for one training iteration per batch. The time needed to execute a training iteration per batch equals the time taken to complete a forward propagation operation followed by a backpropagation operation. Figure \ref{fig-perf-CNNSpeedup} summarizes the training time per iteration per batch for both CPU and GPUs (left y-axis), and the corresponding GPU speedup over CPU (right y-axis). 

These findings for four different CNN network models (i.e., Alexnet \cite{krizhevsky_alexnet_2012}, GoogLeNet \cite{szegedy_googlenet_2015}, Overfeat \cite{sermanet_overfeat_2013}, and VGG \cite{simonyan_vgg_2014}) available in four tools (i.e., Caffe, Tensorflow, Theano, and Torch) \cite{murphy_benchmark_2017} clearly suggest that network training process is much accelerated in GPUs in comparison to CPUs. Moreover, another important message is that, all GPUs are not the same and all tools don't scale up at the same rate. The time required to train a neural network  strongly depends on which DL framework is being used.
As for the hardware platform, the Tesla P100 accelerator provides the best speedup with Tesla M40 being the second and Tesla K80 being the last among the three. In CPUs, TensorFlow achieves the least training time indicating a quicker training of the network. In GPUs, Caffe usually provides the best speedup over CPU but Tensorflow and Torch perform faster training than Cafee. Though Tensorflow and Torch have similar performances (indicated by the height of the lines), Torch slightly outperforming Tensorflow in most of the networks. Finally, most of the tools outperform Theano.

% \section{Case Studies}
% \label{sec-case-stud}
% This section will contain case studies of individual architectures for a specific datatype from any of the available tools.

% \textcolor{red}{Should contain a summary table based on the qualitative and quantitative analysis results discussed in the previous section; and which is more appropriate for which dataset.}

% \subsection{Omics}
% \label{subsec-cs-omics}
% A step-by-step example of any omics data using DA/RNN/DBN.

% \subsection{Medical Imaging}
% \label{subsec-cs-medimg}
% A step-by-step example of analyzing medical images using CNN.

% \subsection{Brain-Machine Interfacing}
% \label{subsec-cs-bmi}
% A step-by-step example of any brain signal using DA/RNN/DBN.

\section{Open Issues and Future Perspectives}
\label{sec-issues-persp}
Brain has the capability to recognize and understand patterns almost instantaneously. Since last several decades, scientists have been trying decode the biological mechanism of natural pattern recognition takes place in the brain and translate that principle in AI systems. The increasing knowledge about the brain's information processing policies enabled this analogy to be adopted and implemented in computing systems. Recent technological breakthroughs, seamless integration of diverse techniques, better understanding of the learning systems, declination of computing costs, and expansion of computational power empowered computing systems to reach human level intelligence in certain scenarios \cite{mnih_human-level_2015}.
Nonetheless, many of these methods require improvements in order not to fall short in situations they fail at present. In this line, we identify below shortcomings and bottlenecks of the popular methods, open research questions and challenges, and outline possible directions which requires attention in the near future.

%Brain solves problems through reinforcement learning and hierarchical sensory processing systems. Though since the 1950's the field of AI has been trying to adopt and implement this analogy in computers, notable progress has been seen only recently. This has been possible due to our increasing understanding about learning systems, increase of computational power, decline of computing costs, and last but not the least, the seamless integration of different technological and technical breakthroughs. However, there are still situations where these methods fail and / or may be improved. Below we outline, what in our opinion are, the lacking of the current techniques, the existing open research challenges, and speculate some future perspectives that will facilitate further advancement of the field. 

%The combined computational capability and flexibility provided by the two prominent ML methods (i.e., DL and RL) also have limitations \cite{ravi_dl_2017}. 
First of all, DL methods usually require large datasets. Though the computing cost is declining with increasing computational power and speed, it is not worthy to apply the DL methods in cases of small to moderate sized datasets. Besides, considering that many of the DL methods perform continuous geometric transformations of one data manifold to another with an assumption that there exist learnable transfer functions which can perform the mapping \cite{chollet_dl_lim_2017}. However, in cases when the relationships among the data are causal or very complex to be learned by the geometric transformations, the DL methods fail regardless the size of the dataset \cite{zenil_causal_reprogramming_2017}. 
%These include situations requiring reasoning-like programming or algorithmic-like data manipulation \cite{chollet_dl_lim_2017}.
Also, interpreting high level outcomes of DL methods are difficult due to inadequate in-depth understanding of the DL theories which causes many of such models to be considered as `Black box' \cite{shwartz-ziv_bb_dnn_2017}.
%\cite{Erhan-vis-techreport-2010}. 
Moreover, like many other ML techniques, DL is also susceptible to misclassification \cite{nguyen_dl_fool_2015} and over-classification \cite{szegedy_ipnn_2014}. 

Additionally, harnessing full benefits offered by the open access data repositories, in terms of data sharing and re-use, are often hampered by the lack of unified reporting data standards and non-uniformity of reported information \cite{baker_standardizing_data_2013}. Data provenance, curation, and annotation of these biological big data is a huge challenge too \cite{wittig_data_2017}.

%Moreover, in representing action-value pairs in RL, it is not possible to use all nonlinear approximators which may cause instability or divergence \cite{mnih_human-level_2015}. Also, bootstrapping makes many of the RL algorithms NP hard and inapplicable to real-time applications as they are too slow and in some cases too dangerous (e.g., autonomous driving). 

Furthermore, except very few large enterprises, the power of distributed and parallel computation through cloud computing remained unexplored for the DL techniques. Due to the fact that the DL techniques require retraining for different datasets, repeated training becomes a bottleneck for cloud computing environments. Also, in such distributed environments, data privacy and security concerns are still prevailing \cite{mahmud_soa_2012}, and real-time processing capability of experimental data is underdeveloped \cite{mahmud_webqst_2014}.

% DL
To mitigate the shortcomings and address the open issues, the existing theoretical foundations of the DL methods need to be improved. The DL models are required not only to be able to describe specific data but also generalize them on the basis of experimental data which is crucial to quantify the performances of individual NN models \cite{angelov_dl_challenges_2016}. These improvements should take place in several directions and address issues like-- quantitative assessment of individual model's learning efficiency and associated computational complexity in relation to well defined parameter tuning strategies, the ability to generalize and topologically self-organize based on data-driven properties. Also, to facilitate intuitive and less cumbersome interpretation of the analysis results, novel tools for data visualization should be incorporated in the DL frameworks. 
% RL
%In terms of learning strategies, updated hybrid on- and off-policy with new advances in optimization techniques are required. The problems pertaining to observability of RL are yet to be completely solved, and optimal action selection is still a huge challenge.

% deep RL
Recent developments in combined methods pertaining to deep reinforcement learning (deep RL) have been popularly applied to many application domains (for a review on deep RL, see \cite{arulkumaran_deep_rl_2017}). However, deep RL methods have not yet been applied to biological pattern recognition problems. For example, analyzing and aggregating dynamically changing patterns in biological data coming from multiple levels could help to remove data redundancy and discover novel biomarkers for disease detection and prevention. Also, novel deep RL methods are needed to reduce the currently required large-set of labeled training data. 

% Data sources
Renewing efforts are required for standardization, annotation, curation, and provenance of data and their sources along with ensuring uniformity of information among the different repositories.
% Tools
Additionally, to keep up with the rapidly growing big data, powerful and secure computational infrastructures in terms of distributed, cloud, and parallel computing tailored to such well-understood learning mechanisms are badly needed.
% Benchmarking
Lastly, there are many other popular DL tools (e.g., Keras, Chainer, Lasagne) and architectures (e.g., DBN) which need to be benchmarked providing the users with a more comprehensive list to choose. Also, the currently available benchmarks are mostly performed on non biological data, and their scalability to biological data aren't very well, thus, specialized benchmarking on biological data are needed. 

\section{Conclusion}
\label{sec-conclusion}
The biological big data coming from different application domains are multimodal, multidimentional, and complex in nature. At present, a great deal of such big data are publicly available. The affordable access to these data came with a huge challenge to analyze patterns in them which require sophisticated ML tools to do the job. As a result, many ML based analytical tools have been developed and reported over the last decades and this process has been facilitated greatly by the decrease of computational costs, increase of computing power, and availability of cheap storage. With the help of these learning techniques, machines have been trained to understand and decipher complex patterns and interactions of variables in biological data. To facilitate a wider dissemination of DL techniques applied to biological big data and serve as a reference point, this article provides a comprehensive survey of the literature on those techniques' application on biological data and the relevant open access data repositories. It also lists existing open source tools and frameworks implementing various DL methods, and compares these tools for their popularity and performance. Finally, it concludes by pointing out some open issues and proposing some future perspectives.

\nolinenumbers

%This defines the bibliographies style. Search online for a list of available styles.
\bibliographystyle{IEEEtran}

%This is where your bibliography is generated. Make sure that your .bib file is actually called library.bib
\bibliography{ref}

\begin{thebibliography}{100}
\providecommand{\url}[1]{#1}
\csname url@samestyle\endcsname
\providecommand{\newblock}{\relax}
\providecommand{\bibinfo}[2]{#2}
\providecommand{\BIBentrySTDinterwordspacing}{\spaceskip=0pt\relax}
\providecommand{\BIBentryALTinterwordstretchfactor}{4}
\providecommand{\BIBentryALTinterwordspacing}{\spaceskip=\fontdimen2\font plus
\BIBentryALTinterwordstretchfactor\fontdimen3\font minus
  \fontdimen4\font\relax}
\providecommand{\BIBforeignlanguage}[2]{{%
\expandafter\ifx\csname l@#1\endcsname\relax
\typeout{** WARNING: IEEEtran.bst: No hyphenation pattern has been}%
\typeout{** loaded for the language `#1'. Using the pattern for}%
\typeout{** the default language instead.}%
\else
\language=\csname l@#1\endcsname
\fi
#2}}
\providecommand{\BIBdecl}{\relax}
\BIBdecl

\bibitem{Coleman_biology_1977}
W.~Coleman, \emph{Biology in the nineteenth century : problems of form,
  function, and transformation}.\hskip 1em plus 0.5em minus 0.4em\relax
  Cambridge ; New York: Cambridge University Press, 1977.

\bibitem{Magner_history_2002}
L.~N. Magner, \emph{A history of the life sciences}, 3rd~ed.\hskip 1em plus
  0.5em minus 0.4em\relax New York: M. Dekker, 2002.

\bibitem{Brenner_history_2012}
S.~Brenner, ``History of science. the revolution in the life sciences,''
  \emph{Science}, vol. 338, no. 6113, pp. 1427--8, 2012.

\bibitem{shendure_next-generation_2008}
J.~Shendure and H.~Ji, ``\BIBforeignlanguage{en}{Next-generation {DNA}
  sequencing},'' \emph{\BIBforeignlanguage{en}{Nat. Biotechnol.}}, vol.~26,
  no.~10, pp. 1135--1145, Oct. 2008.

\bibitem{metzker_sequencing_2010}
M.~L. Metzker, ``\BIBforeignlanguage{en}{Sequencing technologies — the next
  generation},'' \emph{\BIBforeignlanguage{en}{Nat. Rev. Genet.}}, vol.~11,
  no.~1, pp. 31--46, Jan. 2010.

\bibitem{Vadivambal_bioimaging_2016}
R.~Vadivambal and D.~S. Jayas, \emph{Bio-imaging : principles, techniques, and
  applications}.\hskip 1em plus 0.5em minus 0.4em\relax Boca Raton, FL: CRC
  Press, Taylor \& Francis Group, 2016.

\bibitem{poldrack_progress_2015}
R.~A. Poldrack and M.~J. Farah, ``\BIBforeignlanguage{en}{Progress and
  challenges in probing the human brain},''
  \emph{\BIBforeignlanguage{en}{Nature}}, vol. 526, no. 7573, pp. 371--379,
  Oct. 2015.

\bibitem{Lebedev-bmi-2017}
M.~A. Lebedev and M.~A.~L. Nicolelis, ``Brain-machine interfaces: From basic
  science to neuroprostheses and neurorehabilitation,'' \emph{Phys. Rev.},
  vol.~97, no.~2, pp. 767--837, 2017.

\bibitem{quackenbush_extracting_2007}
J.~Quackenbush, ``Extracting biology from high-dimensional biological data,''
  \emph{J. Exp. Biol.}, vol. 210, pp. 1507--17, 2007.

\bibitem{mattmann_computing_2013}
C.~A. Mattmann, ``\BIBforeignlanguage{en}{Computing: {A} vision for data
  science},'' \emph{\BIBforeignlanguage{en}{Nature}}, vol. 493, no. 7433, pp.
  473--475, Jan. 2013.

\bibitem{li_big_2014}
Y.~Li and L.~Chen, ``Big {Biological} {Data}: {Challenges} and
  {Opportunities},'' \emph{Genomics Proteomics Bioinformatics}, vol.~12, no.~5,
  pp. 187--189, Oct. 2014.

\bibitem{marx_biology_2013}
V.~Marx, ``\BIBforeignlanguage{en}{Biology: {The} big challenges of big
  data},'' \emph{\BIBforeignlanguage{en}{Nature}}, vol. 498, no. 7453, pp.
  255--260, Jun. 2013.

\bibitem{tarca_machine_2007}
A.~L. Tarca, V.~J. Carey, X.-w. Chen, R.~Romero, and S.~Draghici, ``Machine
  learning and its applications to biology.'' \emph{PLoS Comput. Biol.},
  vol.~3, no.~6, p. e116, 2007.

\bibitem{cheng_nn_review_1994}
B.~Cheng and D.~Titterington, ``Neural {Networks}: {A} {Review} from a
  {Statistical} {Perspective},'' \emph{Stat. Sc.}, vol.~9, pp. 2--30, 1994.

\bibitem{jain_ann_review_1996}
A.~Jain, J.~Mao, and K.~Mohiuddin, ``Artificial neural networks: a tutorial,''
  \emph{Computer}, vol.~29, no.~3, pp. 31--44, Mar. 1996.

\bibitem{kotsiantis_ml_review_2006}
S.~B. Kotsiantis, I.~D. Zaharakis, and P.~E. Pintelas,
  ``\BIBforeignlanguage{en}{Machine learning: a review of classification and
  combining techniques},'' \emph{\BIBforeignlanguage{en}{Artif. Intell. Rev.}},
  vol.~26, no.~3, pp. 159--190, Nov. 2006.

\bibitem{horgan_omic_2011}
R.~P. Horgan and L.~C. Kenny, ``\BIBforeignlanguage{en}{‘{Omic}’
  technologies: genomics, transcriptomics, proteomics and metabolomics},''
  \emph{\BIBforeignlanguage{en}{Obstet. Gynecol.}}, vol.~13, no.~3, pp.
  189--195, Jul. 2011.

\bibitem{libbrecht_ml_2015}
M.~W. Libbrecht and W.~S. Noble, ``\BIBforeignlanguage{en}{Machine learning
  applications in genetics and genomics},'' \emph{\BIBforeignlanguage{en}{Nat.
  Rev. Genet.}}, vol.~16, no.~6, pp. 321--332, Jun. 2015.

\bibitem{kan_machine_2017}
A.~Kan, ``\BIBforeignlanguage{en}{Machine learning applications in cell image
  analysis},'' \emph{\BIBforeignlanguage{en}{Immunol. Cell. Biol.}}, Mar. 2017.

\bibitem{vidaurre_machine-learning-based_2010}
C.~Vidaurre, C.~Sannelli, K.-R. M{\"u}ller, and B.~Blankertz,
  ``Machine-{Learning}-{Based} {Coadaptive} {Calibration} for
  {Brain}-{Computer} {Interfaces},'' \emph{Neural Computat.}, vol.~23, no.~3,
  pp. 791--816, Dec. 2010.

\bibitem{mala_feature_2014}
S.~Mala and K.~Latha, ``\BIBforeignlanguage{en}{Feature {Selection} in
  {Classification} of {Eye} {Movements} {Using} {Electrooculography} for
  {Activity} {Recognition}},'' \emph{\BIBforeignlanguage{en}{Com. Math. Met.
  Med.}}, vol. 2014, Dec. 2014.

\bibitem{mahmud_processing_2016}
M.~Mahmud and S.~Vassanelli, ``\BIBforeignlanguage{English}{Processing and
  {Analysis} of {Multichannel} {Extracellular} {Neuronal} {Signals}:
  {State}-of-the-{Art} and {Challenges}},''
  \emph{\BIBforeignlanguage{English}{Front. Neurosci.}}, vol.~10, 2016.

\bibitem{lemm_introduction_2011}
S.~Lemm, B.~Blankertz, T.~Dickhaus, and K.-R. M{\"u}ller, ``Introduction to
  machine learning for brain imaging,'' \emph{NeuroImage}, vol.~56, no.~2, pp.
  387--399, May 2011.

\bibitem{erickson_machine_2017}
B.~J. Erickson, P.~Korfiatis, Z.~Akkus, and T.~L. Kline, ``Machine {Learning}
  for {Medical} {Imaging},'' \emph{RadioGraphics}, vol.~37, no.~2, pp.
  505--515, Feb. 2017.

\bibitem{mnih_human-level_2015}
V.~Mnih, K.~Kavukcuoglu, D.~Silver, A.~A. Rusu, J.~Veness, and et~al.,
  ``Human-level control through deep reinforcement learning,'' \emph{Nature},
  vol. 518, no. 7540, pp. 529--533, Feb. 2015.

\bibitem{bengio_learning_2009}
Y.~Bengio, ``Learning deep architectures for ai,'' \emph{Found. Trends Mach.
  Learn.}, vol.~2, no.~1, pp. 1--127, Jan. 2009.

\bibitem{Goodfellow-et-al-2016}
I.~Goodfellow, Y.~Bengio, and A.~Courville, \emph{Deep Learning}.\hskip 1em
  plus 0.5em minus 0.4em\relax Cambridge, USA: MIT Press, 2016.

\bibitem{mahmud_DL_app_2017}
M.~Mahmud, M.~S. Kaiser, A.~Hussain, and S.~Vassanelli, ``Applications of
  {Deep} {Learning} and {Reinforcement} {Learning} to {Biological} {Data},''
  \emph{CoRR}, vol. abs/1711.03985, 2017.

\bibitem{baldi_autoencoder_2012}
P.~Baldi, ``Autoencoders, unsupervised learning and deep architectures,'' in
  \emph{Proc. ICUTLW}, 2012, pp. 37--50.

\bibitem{shen_dl_mia_2017}
D.~Shen, G.~Wu, and H.-I. Suk, ``Deep learning in medical image analysis,''
  \emph{Annu. Rev. Biomed. Eng.}, vol.~19, pp. 221--248, 2017.

\bibitem{bengio_vanishing-gradient_1994}
Y.~Bengio, P.~Simard, and P.~Frasconi, ``Learning long-term dependencies with
  gradient descent is difficult,'' \emph{IEEE Trans. Neural Netw.}, vol.~5,
  no.~2, pp. 157--166, 1994.

\bibitem{vincent_denoising_autoencoders_2010}
P.~Vincent, H.~Larochelle, I.~Lajoie, Y.~Bengio, and P.-A. Manzagol, ``Stacked
  denoising autoencoders: Learning useful representations in a deep network
  with a local denoising criterion,'' \emph{J. M. L. Res.}, vol.~11, pp.
  3371--3408, Dec. 2010.

\bibitem{ranzato_sparse_autoencoders_2006}
M.~Ranzato, C.~Poultney, S.~Chopra, and Y.~LeCun, ``Efficient learning of
  sparse representations with an energy-based model,'' in \emph{Proc. NIPS},
  2006, pp. 1137--1144.

\bibitem{kingma_variational_auto_encoding_2014}
D.~P. Kingma and M.~Welling, ``Auto-{Encoding} {Variational} {Bayes},''
  \emph{CoRR}, Dec. 2014, coRR: 1312.6114.

\bibitem{rifai_contractingauto_encoders_2011}
S.~Rifai, P.~Vincent, X.~Muller, X.~Glorot, and Y.~Bengio, ``Contracting
  auto-encoders: Explicit invariance during feature extraction,'' in
  \emph{Proc. ICML2011}, Bellevue, WA, USA, 2011.

\bibitem{Salakhutdinov_dbm_2009}
R.~Salakhutdinov and G.~E. Hinton, ``Deep boltzmann machines,'' in \emph{Proc.
  AISTATS2009}, 2009, pp. 448--455.

\bibitem{zhou_dl_mia_book_2017}
S.~K. Zhou, H.~Greenspan, and D.~Shen, \emph{Deep Learning for Medical Image
  Analysis}.\hskip 1em plus 0.5em minus 0.4em\relax Academic Press, 2017.

\bibitem{geman_gibbs_sampling_1984}
S.~Geman and D.~Geman, ``Stochastic relaxation, gibbs distributions, and the
  bayesian restoration of images,'' \emph{IEEE Trans. Pattern Anal. Mach.
  Intell.}, vol.~6, no.~6, pp. 721--741, 1984.

\bibitem{hinton_dbn_2006}
G.~E. Hinton, S.~Osindero, and Y.-W. Teh, ``A fast learning algorithm for deep
  belief nets,'' \emph{Neural Comput.}, vol.~18, no.~7, pp. 1527--1554, Jul.
  2006.

\bibitem{ravi_dl_2017}
D.~Ravi, C.~Wong, F.~Deligianni, M.~Berthelot, and et~al., ``Deep {Learning}
  for {Health} {Informatics},'' \emph{IEEE J. Biomed. Health Inform.}, vol.~21,
  no.~1, pp. 4--21, Jan. 2017.

\bibitem{lecun_cnn_1998}
Y.~LeCun and Y.~Bengio, ``The handbook of brain theory and neural networks,''
  in \emph{The Handbook of Brain Theory and Neural Networks}, M.~A. Arbib,
  Ed.\hskip 1em plus 0.5em minus 0.4em\relax Cambridge, MA, USA: MIT Press,
  1998, ch. Convolutional Networks for Images, Speech, and Time Series, pp.
  255--258.

\bibitem{bouvrie_notes_2006}
\BIBentryALTinterwordspacing
J.~Bouvrie, ``Notes on {Convolutional} {Neural} {Networks},'' 2006. [Online].
  Available: \url{http://cogprints.org/5869/}
\BIBentrySTDinterwordspacing

\bibitem{krizhevsky_alexnet_2012}
A.~Krizhevsky, I.~Sutskever, and G.~E. Hinton, ``{ImageNet} classification with
  deep convolutional neural networks,'' in \emph{Proc. NIPS}, 2012, pp.
  1097--1105.

\bibitem{simonyan_vgg_2014}
K.~Simonyan and A.~Zisserman, ``Very deep convolutional networks for
  large-scale image recognition,'' \emph{CoRR}, vol. abs/1409.1556, 2014.

\bibitem{szegedy_googlenet_2015}
C.~Szegedy, W.~Liu, Y.~Jia, P.~Sermanet, S.~Reed, D.~Anguelov, D.~Erhan,
  V.~Vanhoucke, and A.~Rabinovich, ``Going deeper with convolutions,'' in
  \emph{Proc. CVPR2015}, 2015, pp. 1--9.

\bibitem{elman_finding_1990}
J.~L. Elman, ``\BIBforeignlanguage{en}{Finding {Structure} in {Time}},''
  \emph{\BIBforeignlanguage{en}{Cognitive Sci.}}, vol.~14, no.~2, pp. 179--211,
  Mar. 1990.

\bibitem{zeiler_relu_2013}
M.~D. Zeiler, M.~Ranzato, R.~Monga, M.~Mao, K.~Yang, Q.~V. Le, P.~Nguyen,
  A.~Senior, V.~Vanhoucke, J.~Dean, and G.~E. Hinton, ``On rectified linear
  units for speech processing,'' in \emph{Proc. ICASSP}, 2013, pp. 3517--3521.

\bibitem{lipton_critical_2015}
Z.~C. Lipton, J.~Berkowitz, and C.~Elkan, ``A {Critical} {Review} of
  {Recurrent} {Neural} {Networks} for {Sequence} {Learning},'' \emph{CoRR}, May
  2015, coRR: 1506.00019.

\bibitem{schuster_bidirectional_rnn_1997}
M.~Schuster and K.~K. Paliwal, ``Bidirectional recurrent neural networks,''
  \emph{IEEE Tran. Signal Proces.}, vol.~45, no.~11, pp. 2673--2681, Nov. 1997.

\bibitem{hochreiter_lstm_1997}
S.~Hochreiter and J.~Schmidhuber, ``Long short-term memory,'' \emph{Neural
  Comput.}, vol.~9, no.~8, pp. 1735--1780, Nov. 1997.

\bibitem{cho_grnn_2014}
K.~Cho, B.~van Merrienboer, {\c{C}}.~G{\"{u}}l{\c{c}}ehre, F.~Bougares,
  H.~Schwenk, and Y.~Bengio, ``Learning phrase representations using {RNN}
  encoder-decoder for statistical machine translation,'' in \emph{Proc. EMNLP},
  2014, pp. 1724--1734.

\bibitem{litjens_mia_2017}
G.~Litjens, T.~Kooi, B.~E. Bejnordi, A.~A.~A. Setio, F.~Ciompi, M.~Ghafoorian,
  J.~A. W.~M. van~der Laak, B.~van Ginneken, and C.~I. Sánchez, ``A survey on
  deep learning in medical image analysis,'' \emph{Med. Image Anal.}, vol.~42,
  pp. 60--88, 2017.

\bibitem{danaee2016}
P.~Danaee, R.~Ghaeini, and D.~A. Hendrix, ``A deep learning approach for cancer
  detection and relevant gene identification,'' in \emph{Proc. Pac. Symp.
  Biocomput.}, vol.~22, 2016, pp. 219--229.

\bibitem{li_2016}
H.~Li, ``A template-based protein structure reconstruction method using da
  learning,'' \emph{J. Proteomics Bioinform.}, vol.~9, no.~12, 2016.

\bibitem{Lee3045382}
T.~Lee and S.~Yoon, ``Boosted categorical restricted boltzmann machine for
  computational prediction of splice junctions,'' in \emph{Proc. ICML}, 2015,
  pp. 2483--2492.

\bibitem{6944490}
R.~Ibrahim, N.~A. Yousri, M.~A. Ismail, and N.~M. El-Makky, ``Multi-level
  gene/mirna feature selection using deep belief nets and active learning,'' in
  \emph{Proc. IEEE EMBC}, Aug 2014, pp. 3957--3960.

\bibitem{Chen2015}
L.~Chen, C.~Cai, V.~Chen, and X.~Lu, ``Trans-species learning of cellular
  signaling systems with bimodal deep belief networks,'' \emph{Bioinformatics},
  vol.~31, no.~18, pp. 3008--3015, sep 2015.

\bibitem{pan_rbm_2017}
X.~Pan and H.-B. Shen, ``{RNA}-protein binding motifs mining with a new hybrid
  deep learning based cross-domain knowledge integration approach,'' \emph{BMC
  Bioinform.}, vol.~18, no.~1, 2017.

\bibitem{Denas2013DeepMO}
O.~Denas and J.~Taylor, ``Deep modeling of gene expression regulation in
  erythropoiesis model,'' in \emph{Proc. ICMLRL}, 2013.

\bibitem{Kelley2016}
D.~R. Kelley, J.~Snoek, and J.~L. Rinn, ``Basset: learning the regulatory code
  of the accessible genome with deep convolutional neural networks,''
  \emph{Genome Res.}, vol.~26, no.~7, pp. 990--9, 2016.

\bibitem{DBLPZengELG16}
H.~Zeng, M.~D. Edwards, G.~Liu, and D.~K. Gifford, ``Convolutional neural
  network architectures for predicting dna-protein binding,''
  \emph{Bioinformatics}, vol.~32, no.~12, pp. 121--127, 2016.

\bibitem{citeulike13721890}
J.~Zhou and O.~G. Troyanskaya, ``{Predicting effects of noncoding variants with
  deep learning-based sequence model},'' \emph{Nature Methods}, vol.~12,
  no.~10, pp. 931--934, Aug. 2015.

\bibitem{Huang069682}
Y.~Huang, B.~Gulko, and A.~Siepel, ``Fast, scalable prediction of deleterious
  noncoding variants from functional and population genomic data,''
  \emph{Nature Genet}, vol.~49, pp. 618--624, 2017.

\bibitem{DBLPWang0MX15}
S.~Wang, J.~Peng, J.~Ma, and J.~Xu, ``Protein secondary structure prediction
  using deep convolutional neural fields,'' \emph{Scientific Reports}, vol.~6,
  no.~1, Nov 2016.

\bibitem{alipanahi_deepbind_2015}
B.~Alipanahi, A.~Delong, M.~T. Weirauch, and B.~J. Frey, ``Predicting the
  sequence specificities of dna- and rna-binding proteins by deep learning,''
  \emph{Nature Biotechnol.}, vol.~33, no.~8, pp. 831--838, 2015.

\bibitem{DBLPParkMCY16}
S.~Park, S.~Min, H.~Choi, and S.~Yoon, ``{deepMiRGene}: Deep neural network
  based precursor microrna prediction,'' \emph{CoRR}, vol. abs/1605.00017,
  2016.

\bibitem{DBLPLeeBPY16}
B.~Lee, J.~Baek, S.~Park, and S.~Yoon, ``{deepTarget}: End-to-end learning
  framework for {miRNA} target prediction using deep recurrent neural
  networks,'' \emph{CoRR}, vol. abs/1603.09123, 2016.

\bibitem{Ciresan2013}
D.~Ciresan, A.~Giusti, L.~Gambardella, and J.~Schmidhuber, ``Mitosis detection
  in breast cancer histology images with deep neural networks,'' in \emph{Proc.
  MICCAI}, 2013, pp. 411--4188.

\bibitem{NIPS2012_4741}
------, ``Deep neural nets segment neuronal membrane in electron microscopy
  images,'' in \emph{Proc. NIPS}, 2012, pp. 2843--2851.

\bibitem{DBLPStollengaBLS15}
M.~F. Stollenga, W.~Byeon, M.~Liwicki, and J.~Schmidhuber, ``Parallel
  multi-dimensional lstm, with application to fast biomedical volumetric image
  segmentation,'' in \emph{Proc. NIPS}, 2015, pp. 2980--88.

\bibitem{DBLPHavaeiGLJ16}
M.~Havaei, N.~Guizard, H.~Larochelle, and P.-M. Jodoin, \emph{Deep Learning
  Trends for Focal Brain Pathology Segmentation in MRI}.\hskip 1em plus 0.5em
  minus 0.4em\relax Cham: Springer, 2016, pp. 125--148.

\bibitem{DBLPHosseiniAslGE16}
E.~HosseiniAsl, G.~L. Gimelfarb, and A.~El{-}Baz, ``Alzheimer's disease
  diagnostics by a deeply supervised adaptable 3d convolutional network,''
  \emph{CoRR}, vol. abs/1607.00556, 2016.

\bibitem{Suk2014569}
H.-I. Suk, S.-W. Lee, and D.~Shen, ``Hierarchical feature representation and
  multimodal fusion with deep learning for ad/mci diagnosis,''
  \emph{NeuroImage}, vol. 101, pp. 569 -- 582, 2014.

\bibitem{Li2014}
F.~Li, L.~Tran, K.~H. Thung, S.~Ji, D.~Shen, and J.~Li, ``A {Robust} {Deep}
  {Model} for {Improved} {Classification} of {AD}/{MCI} {Patients},''
  \emph{IEEE J. Biomed. Health. Inform.}, vol.~19, no.~5, pp. 1610--1616, Sep.
  2015.

\bibitem{Kleesiek2016460}
J.~Kleesiek, G.~Urban, A.~Hubert, D.~Schwarz, K.~Maier-Hein, M.~Bendszus, and
  A.~Biller, ``Deep {MRI} brain extraction: A {3D} convolutional neural network
  for skull stripping,'' \emph{NeuroImage}, vol. 129, pp. 460 -- 469, 2016.

\bibitem{kamnitsas_3dcnn_2017}
K.~Kamnitsas, C.~Ledig, V.~F. Newcombe, J.~Simpson, and {et al.}, ``Efficient
  multi-scale 3d {CNN} with fully connected {CRF} for accurate brain lesion
  segmentation,'' \emph{Med. Image Anal.}, vol.~36, pp. 61--78, 2017.

\bibitem{Fritscher2016}
K.~Fritscher, P.~Raudaschl, P.~Zaffino, M.~F. Spadea, G.~C. Sharp, and et~al.,
  ``Deep neural networks for fast segmentation of 3d medical images,'' in
  \emph{Proc. MICCAI}, 2016, pp. 158--165.

\bibitem{DBLPChoLSCD15}
J.~Cho, K.~Lee, E.~Shin, G.~Choy, and S.~Do, ``Medical image deep learning with
  hospital pacs dataset,'' \emph{CoRR}, vol. abs/1511.06348, 2015.

\bibitem{7062868}
D.~Kuang and L.~He, ``Classification on adhd with deep learning,'' in
  \emph{Proc. CCBD}, 2014, pp. 27--32.

\bibitem{ngo_dl_hrt_2017}
T.~Ngo and et~al., ``Combining deep learning and level set for the automated
  segmentation of the left ventricle of the heart from cardiac cine mr,''
  \emph{Med. Image Anal.}, vol.~35, pp. 159--171, 2017.

\bibitem{jirayucharoensak2014}
S.~Jirayucharoensak, S.~Pan-Ngum, and P.~Israsena, ``Eeg-based emotion
  recognition using deep learning network with principal component based
  covariate shift adaptation,'' \emph{Scientific World J.}, pp. 1--10, 2014.

\bibitem{lu_rbm_mi_2016}
N.~Lu, T.~Li, X.~Ren, and H.~Miao, ``A {Deep} {Learning} {Scheme} for {Motor}
  {Imagery} {Classification} based on {Restricted} {Boltzmann} {Machines},''
  \emph{IEEE Trans. Neural Syst. Rehabil. Eng.}, vol.~PP, no.~99, pp. 1--1,
  2016.

\bibitem{yang2015}
H.~Yang, S.~Sakhavi, K.~K. Ang, and C.~Guan, ``On the use of convolutional
  neural networks and augmented csp features for multi-class motor imagery of
  eeg signals classification,'' in \emph{Proc. 37th IEEE EMBC}, 2015, pp.
  2620--2623.

\bibitem{tabar_cnn_mi_eeg_2017}
Y.~R. Tabar and U.~Halici, ``\BIBforeignlanguage{en}{A novel deep learning
  approach for classification of {EEG} motor imagery signals},''
  \emph{\BIBforeignlanguage{en}{J. Neural Eng.}}, vol.~14, no.~1, p. 016003,
  2017.

\bibitem{sakhavi_mi_2015}
S.~Sakhavi, C.~Guan, and S.~Yan, ``\BIBforeignlanguage{en}{Parallel
  convolutional-linear neural network for motor imagery classification},'' in
  \emph{\BIBforeignlanguage{en}{Proc. EUSIPCO}}, 2015, pp. 2786--2790.

\bibitem{li_dbn_as_2013}
K.~Li, X.~Li, Y.~Zhang, and A.~Zhang, ``Affective state recognition from eeg
  with deep belief networks,'' in \emph{Proc. BIBM}, 2013, pp. 305--310.

\bibitem{7033556}
X.~Jia, K.~Li, X.~Li, and A.~Zhang, ``A novel semi-supervised deep learning
  framework for affective state recognition on eeg signals,'' in \emph{Proc.
  IEEE BIBE}, 2014, pp. 30--37.

\bibitem{IAAI1715007}
S.~Tripathi, S.~Acharya, R.~Sharma, S.~Mittal, and et~al., ``Using deep and
  convolutional neural networks for accurate emotion classification on deap
  dataset,'' in \emph{Proc. 29th IAAI}, 2017.

\bibitem{Mirowski20091927}
P.~Mirowski, D.~Madhavan, Y.~LeCun, and R.~Kuzniecky, ``Classification of
  patterns of {EEG} synchronization for seizure prediction,'' \emph{Clin.
  Neurophysiol.}, vol. 120, no.~11, pp. 1927 -- 1940, 2009.

\bibitem{7023547}
M.~Huanhuan and Z.~Yue, ``Classification of electrocardiogram signals with
  dbn,'' in \emph{Proc. IEEE CSE}, 2014, pp. 7--12.

\bibitem{DBLPAtzoriCM16}
M.~Atzori, M.~Cognolato, and H.~M{\"{u}}ller, ``Deep learning with
  convolutional neural networks applied to electromyography data: {A} resource
  for the classification of movements for prosthetic hands,'' \emph{Front.
  Neurorobot.}, vol.~10, p.~9, 2016.

\bibitem{wu_ecg_dbn_2016}
Z.~Wu, X.~Ding, and G.~Zhang, ``A novel method for classification of ecg
  arrhythmias using deep belief networks,'' \emph{J. Comp. Intel. Appl.},
  vol.~15, p. 1650021, 2016.

\bibitem{DBLPYanQWZ0W15}
Y.~Yan, X.~Qin, Y.~Wu, N.~Zhang, J.~Fan, and et~al., ``A restricted boltzmann
  machine based two-lead electrocardiography classification,'' in \emph{Proc.
  BSN}, 2015, pp. 1--9.

\bibitem{GE5}
\BIBentryALTinterwordspacing
``Uci molecular biology (uci mb) protein secondary structure data set,''
  (Accessed on: 17/12/2017). [Online]. Available:
  \url{https://archive.ics.uci.edu/ml/datasets/Molecular+Biology+(Protein+Secondary+Structure)}
\BIBentrySTDinterwordspacing

\bibitem{GE6}
\BIBentryALTinterwordspacing
``Uci molecular biology (uci mb) splice-junction gene sequences data set,''
  (Accessed on: 17/12/2017). [Online]. Available:
  \url{https://archive.ics.uci.edu/ml/datasets/Molecular+Biology+(Splice-junction+Gene+Sequences)}
\BIBentrySTDinterwordspacing

\bibitem{GE7}
\BIBentryALTinterwordspacing
``Uci molecular biology (uci mb) promoter gene sequences data set,'' (Accessed
  on: 17/12/2017). [Online]. Available:
  \url{https://archive.ics.uci.edu/ml/support/Molecular+Biology+(Promoter+Gene+Sequences)}
\BIBentrySTDinterwordspacing

\bibitem{GE11}
\BIBentryALTinterwordspacing
``The international nucleotide sequence database collaboration,'' (Accessed on:
  17/12/2017). [Online]. Available: \url{http://www.insdc.org/}
\BIBentrySTDinterwordspacing

\bibitem{GE11-1}
\BIBentryALTinterwordspacing
``Dna databank of japan,'' (Accessed on: 17/12/2017). [Online]. Available:
  \url{http://www.ddbj.nig.ac.jp/}
\BIBentrySTDinterwordspacing

\bibitem{GE11-2}
\BIBentryALTinterwordspacing
``European nucleotide archive,'' (Accessed on: 17/12/2017). [Online].
  Available: \url{http://www.ebi.ac.uk/ena}
\BIBentrySTDinterwordspacing

\bibitem{GE11-3}
\BIBentryALTinterwordspacing
``Genbank,'' (Accessed on: 17/12/2017). [Online]. Available:
  \url{https://www.ncbi.nlm.nih.gov/genbank/}
\BIBentrySTDinterwordspacing

\bibitem{GE1}
\BIBentryALTinterwordspacing
``Saccharomyces,'' (Accessed on: 17/12/2017). [Online]. Available:
  \url{https://www.yeastgenome.org/}
\BIBentrySTDinterwordspacing

\bibitem{GE2}
\BIBentryALTinterwordspacing
``Pubchem,'' (Accessed on: 17/12/2017). [Online]. Available:
  \url{https://pubchem.ncbi.nlm.nih.gov/sources/}
\BIBentrySTDinterwordspacing

\bibitem{GE3}
\BIBentryALTinterwordspacing
``Encyclopedia of dna elements,'' (Accessed on: 17/12/2017). [Online].
  Available: \url{https://genome.ucsc.edu/ENCODE/}
\BIBentrySTDinterwordspacing

\bibitem{noauthor_uci_nodate}
\BIBentryALTinterwordspacing
``{UCI} {Machine} {Learning} {Repository}: {Data} {Sets},'' (Accessed on:
  17/12/2017). [Online]. Available:
  \url{https://archive.ics.uci.edu/ml/datasets.html}
\BIBentrySTDinterwordspacing

\bibitem{noauthor_international_nodate}
\BIBentryALTinterwordspacing
``International nucleotide sequence database collaboration,'' (Accessed on:
  17/12/2017). [Online]. Available: \url{http://www.insdc.org/}
\BIBentrySTDinterwordspacing

\bibitem{GE12}
\BIBentryALTinterwordspacing
``Nature scientific data,'' (Accessed on: 17/12/2017). [Online]. Available:
  \url{http://go.nature.com/2g6E1Vm}
\BIBentrySTDinterwordspacing

\bibitem{GE13}
\BIBentryALTinterwordspacing
``Small molecule pathway database,'' (Accessed on: 17/12/2017). [Online].
  Available: \url{http://smpdb.ca/}
\BIBentrySTDinterwordspacing

\bibitem{noauthor_cancer_nodate}
\BIBentryALTinterwordspacing
``The {Cancer} {Genome} {Atlas} {Home} {Page},'' (Accessed on: 17/12/2017).
  [Online]. Available: \url{https://cancergenome.nih.gov/}
\BIBentrySTDinterwordspacing

\bibitem{noauthor_rcsb_nodate}
\BIBentryALTinterwordspacing
``{RCSB} {Protein} {Data} {Bank} - {RCSB} {PDB},'' (Accessed on: 17/12/2017).
  [Online]. Available:
  \url{https://www.rcsb.org/pdb/home/home.do#Category-download}
\BIBentrySTDinterwordspacing

\bibitem{noauthor_gems:_nodate}
\BIBentryALTinterwordspacing
``{GEMS}: {Gene} {Expression} {Model} {Selector},'' (Accessed on: 17/12/2017).
  [Online]. Available: \url{http://www.gems-system.org/}
\BIBentrySTDinterwordspacing

\bibitem{noauthor_cancerds_nodate}
\BIBentryALTinterwordspacing
``Cancer {Program} {Legacy} {Publication} {Resources},'' (Accessed on:
  17/12/2017). [Online]. Available:
  \url{http://portals.broadinstitute.org/cgi-bin/cancer/datasets.cgi}
\BIBentrySTDinterwordspacing

\bibitem{noauthor_bioinformatics_nodate}
\BIBentryALTinterwordspacing
``Bioinformatics {Laboratory},'' (Accessed on: 17/12/2017). [Online].
  Available: \url{http://www.biolab.si/supp/bi-cancer/projections/}
\BIBentrySTDinterwordspacing

\bibitem{mstrazar_ionmf:_2017}
\BIBentryALTinterwordspacing
mstrazar, ``{iONMF}: {Integrative} orthogonal non-negative matrix
  factorization,'' 2017, (Accessed on: 17/12/2017). [Online]. Available:
  \url{https://github.com/mstrazar/iONMF}
\BIBentrySTDinterwordspacing

\bibitem{noauthor_jaspar_nodate}
\BIBentryALTinterwordspacing
``{JASPAR} 2018: {An} open-access database of transcription factor binding
  profiles,'' (Accessed on: 17/12/2017). [Online]. Available:
  \url{http://jaspar.genereg.net}
\BIBentrySTDinterwordspacing

\bibitem{noauthor_sysgensim_nodate}
\BIBentryALTinterwordspacing
``{SysGenSIM} - {Benchmark} datasets,'' (Accessed on: 17/12/2017). [Online].
  Available: \url{http://sysgensim.sourceforge.net/datasets.html}
\BIBentrySTDinterwordspacing

\bibitem{noauthor_mirboost_nodate}
\BIBentryALTinterwordspacing
``{miRBoost},'' (Accessed on: 17/12/2017). [Online]. Available:
  \url{https://evryrna.ibisc.univ-evry.fr/evryrna/mirboost/mirboost_help}
\BIBentrySTDinterwordspacing

\bibitem{GE14}
\BIBentryALTinterwordspacing
``Indian genetic disease database,'' (Accessed on: 17/12/2017). [Online].
  Available: \url{www.igdd.iicb.res.in/}
\BIBentrySTDinterwordspacing

\bibitem{Bio2}
\BIBentryALTinterwordspacing
``Image science,'' (Accessed on: 17/12/2017). [Online]. Available:
  \url{https://imagescience.org/}
\BIBentrySTDinterwordspacing

\bibitem{Bio3}
\BIBentryALTinterwordspacing
``Bioimaging,'' (Accessed on: 12/02/2017). [Online]. Available:
  \url{https://omictools.com/image-data-category}
\BIBentrySTDinterwordspacing

\bibitem{Bio4}
\BIBentryALTinterwordspacing
``Cell image library,'' (Accessed on: 17/12/2017). [Online]. Available:
  \url{http://www.cellimagelibrary.org/}
\BIBentrySTDinterwordspacing

\bibitem{Bio5}
\BIBentryALTinterwordspacing
``Bdtnp bio imaging meta database,'' {Requires} {Login}. (Accessed on:
  17/12/2017). [Online]. Available: \url{http://bdtnp.lbl.gov/BID/}
\BIBentrySTDinterwordspacing

\bibitem{Bio6}
\BIBentryALTinterwordspacing
``Eurobioimaging,'' (Accessed on: 17/12/2017). [Online]. Available:
  \url{http://www.eurobioimaging.eu/}
\BIBentrySTDinterwordspacing

\bibitem{Bio7}
\BIBentryALTinterwordspacing
``The cell,'' now part of cell image library. (Accessed on: 12/02/2017).
  [Online]. Available: \url{http://ccdb.ucsd.edu/index.shtm}
\BIBentrySTDinterwordspacing

\bibitem{Bio8}
\BIBentryALTinterwordspacing
``The journal of cell biology data viewer,'' (Accessed on: 17/12/ 2017).
  [Online]. Available: \url{http://jcb-dataviewer.rupress.org/}
\BIBentrySTDinterwordspacing

\bibitem{noauthor_mitos-atypia-14_nodate}
\BIBentryALTinterwordspacing
``{MITOS}-{ATYPIA}-14 - {Dataset},'' (Accessed on: 17/12/2017). [Online].
  Available: \url{https://mitos-atypia-14.grand-challenge.org/dataset/}
\BIBentrySTDinterwordspacing

\bibitem{noauthor_nitrc:_nodate}
\BIBentryALTinterwordspacing
``{NITRC}: {IBSR}: {Tool}/{Resource} {Info},'' (Accessed on: 17/12/2017).
  [Online]. Available: \url{https://www.nitrc.org/projects/ibsr}
\BIBentrySTDinterwordspacing

\bibitem{shattuck_construction_2008}
\BIBentryALTinterwordspacing
D.~W. Shattuck, M.~Mirza, V.~Adisetiyo, C.~Hojatkashani, G.~Salamon, and
  et~al., ``\BIBforeignlanguage{en}{Construction of a 3d probabilistic atlas of
  human cortical structures},'' \emph{\BIBforeignlanguage{en}{NeuroImage}},
  vol.~39, no.~3, pp. 1064--1080, Feb. 2008. [Online]. Available:
  \url{http://linkinghub.elsevier.com/retrieve/pii/S1053811907008099}
\BIBentrySTDinterwordspacing

\bibitem{noauthor_adhd200_nodate}
\BIBentryALTinterwordspacing
``{ADHD}200,'' (Accessed on: 17/12/2017). [Online]. Available:
  \url{http://fcon_1000.projects.nitrc.org/indi/adhd200/}
\BIBentrySTDinterwordspacing

\bibitem{NM2}
\BIBentryALTinterwordspacing
``Open fmri: A multi-subject, multi-modal human neuroimaging dataset,''
  (Accessed on: 17/12/2017). [Online]. Available: \url{https://openfmri.org/}
\BIBentrySTDinterwordspacing

\bibitem{NM3}
\BIBentryALTinterwordspacing
``Open access series of imaging studies (oasis),'' (Accessed on: 17/12/2017).
  [Online]. Available: \url{http://www.oasis-brains.org/}
\BIBentrySTDinterwordspacing

\bibitem{NM4}
\BIBentryALTinterwordspacing
``Neurosynth,'' (Accessed on: 17/12/2017). [Online]. Available:
  \url{http://neurosynth.org/}
\BIBentrySTDinterwordspacing

\bibitem{NM5}
\BIBentryALTinterwordspacing
``Neuroimaging dataset of brain tumour patients,'' (Accessed on: 17/12/2017).
  [Online]. Available: \url{https://goo.gl/fmYYm4}
\BIBentrySTDinterwordspacing

\bibitem{NM6}
\BIBentryALTinterwordspacing
``Autism brain imaging data exchange,'' (Accessed on: 17/12/2017). [Online].
  Available: \url{https://goo.gl/n694sN}
\BIBentrySTDinterwordspacing

\bibitem{NM7}
\BIBentryALTinterwordspacing
``Open neuroimaging datasets,'' (Accessed on: 17/12/2017). [Online]. Available:
  \url{https://goo.gl/azm4XW}
\BIBentrySTDinterwordspacing

\bibitem{Nm8}
\BIBentryALTinterwordspacing
``Neuroimaging informatics tools and resources clearinghouse dataset,''
  (Accessed on: 17/12/2017). [Online]. Available: \url{https://goo.gl/CA2pkO}
\BIBentrySTDinterwordspacing

\bibitem{NM9}
\BIBentryALTinterwordspacing
``Alzheimer's disease neuroimaging initiative (adni datasets,'' (Accessed on:
  17/12/2017). [Online]. Available: \url{http://adni.loni.usc.edu/}
\BIBentrySTDinterwordspacing

\bibitem{NM10}
\BIBentryALTinterwordspacing
``Brain development datasets,'' (Accessed on: 17/12/2017). [Online]. Available:
  \url{http://brain-development.org/ixi-dataset/}
\BIBentrySTDinterwordspacing

\bibitem{NM11}
\BIBentryALTinterwordspacing
``Neurovault,'' (Accessed on: 17/12/2017). [Online]. Available:
  \url{http://neurovault.org/}
\BIBentrySTDinterwordspacing

\bibitem{NM12}
\BIBentryALTinterwordspacing
``The cancer imaging archive,'' (Accessed on: 17/12/2017). [Online]. Available:
  \url{http://www.cancerimagingarchive.net/}
\BIBentrySTDinterwordspacing

\bibitem{BCI01}
B.~B. et~al., ``Bci competition datasets,''
  \url{http://www.bbci.de/competition/ii/}, 2002.

\bibitem{BMI5}
\BIBentryALTinterwordspacing
``Database for emotion analysis using physiological signals,'' (Accessed on:
  17/12/2017). [Online]. Available:
  \url{http://www.eecs.qmul.ac.uk/mmv/datasets/deap/}
\BIBentrySTDinterwordspacing

\bibitem{noauthor_ninapro_nodate}
\BIBentryALTinterwordspacing
``{NinaPro} {Database} — {Non}-{Invasive} {Adaptive} {Hand} {Prosthetics},''
  (Accessed on: 17/12/2017). [Online]. Available:
  \url{https://www.idiap.ch/project/ninapro/database}
\BIBentrySTDinterwordspacing

\bibitem{BMI1}
\BIBentryALTinterwordspacing
``Uci ml repository,'' (Accessed on: 17/12/2017). [Online]. Available:
  \url{https://archive.ics.uci.edu/ml/datasets.html}
\BIBentrySTDinterwordspacing

\bibitem{BMI2}
\BIBentryALTinterwordspacing
``Physionet,'' (Accessed on: 17/12/2017). [Online]. Available:
  \url{https://physionet.org/physiobank/database/}
\BIBentrySTDinterwordspacing

\bibitem{BMI3}
\BIBentryALTinterwordspacing
``Bncihorizon2020,'' (Accessed on: 17/12/2017). [Online]. Available:
  \url{https://goo.gl/6gLj52}
\BIBentrySTDinterwordspacing

\bibitem{BMI4}
\BIBentryALTinterwordspacing
``Mahnob-hci,'' (Accessed on: 17/12/2017). [Online]. Available:
  \url{https://mahnob-db.eu/hci-tagging/}
\BIBentrySTDinterwordspacing

\bibitem{BMI6}
\BIBentryALTinterwordspacing
``Meg-based multimodal database for decoding affective physiological
  responses,'' (Accessed on: 17/12/2017). [Online]. Available:
  \url{http://mhug.disi.unitn.it/wp-content/DECAF/DECAF.html}
\BIBentrySTDinterwordspacing

\bibitem{BMI7}
\BIBentryALTinterwordspacing
``Brain signals data,'' (Accessed on: 17/12/2017). [Online]. Available:
  \url{http://www.brainsignals.de/}
\BIBentrySTDinterwordspacing

\bibitem{BMI8}
\BIBentryALTinterwordspacing
``Tele ecg,'' (Accessed on: 17/12/2017). [Online]. Available:
  \url{https://dataverse.harvard.edu/dataset.xhtml?persistentId=doi:10.7910/DVN/QTG0EP}
\BIBentrySTDinterwordspacing

\bibitem{BMI9}
\BIBentryALTinterwordspacing
R.~Guillaume, ``{LIMO} {EEG} {Dataset},'' 2016, (Accessed on: 17/12/2017).
  [Online]. Available: \url{https://datashare.is.ed.ac.uk/handle/10283/2189}
\BIBentrySTDinterwordspacing

\bibitem{BMI10}
\BIBentryALTinterwordspacing
``Eeg single subject mismatch negativity (essmn),'' (Accessed on: 17/12/2017).
  [Online]. Available: \url{http://bit.ly/2nkLy6X}
\BIBentrySTDinterwordspacing

\bibitem{BMI11}
\BIBentryALTinterwordspacing
``Eeg dataset,'' (Accessed on: 17/12/2017). [Online]. Available:
  \url{http://bit.ly/2mEmZVo}
\BIBentrySTDinterwordspacing

\bibitem{BMI12}
\BIBentryALTinterwordspacing
``Facial s-emg dataset,'' (Accessed on: 17/12/2017). [Online]. Available:
  \url{http://bit.ly/2npCAH8}
\BIBentrySTDinterwordspacing

\bibitem{karpathy_peek_2017}
\BIBentryALTinterwordspacing
A.~Karpathy, ``A {Peek} at {Trends} in {Machine} {Learning},'' Apr. 2017.
  [Online]. Available:
  \url{https://medium.com/@karpathy/a-peek-at-trends-in-machine-learning-ab8a1085a106}
\BIBentrySTDinterwordspacing

\bibitem{bahrampour_dl_fws_2016}
S.~Bahrampour, N.~Ramakrishnan, L.~Schott, and M.~Shah, ``Comparative study of
  deep learning software frameworks,'' \emph{CoRR}, vol. abs/1511.06435, 2016,
  arXiv: 1511.06435.

\bibitem{shi_benchmarking_dl_2016}
S.~Shi and {et al.}, ``Benchmarking state-of-the-art deep learning software
  tools,'' \emph{CoRR}, vol. abs/1608.07249, 2016.

\bibitem{deepmark_benchmark_2017}
\BIBentryALTinterwordspacing
``The deep learning benchmarks,'' 2017, (Accessed on: 17- Dec- 2017). [Online].
  Available: \url{https://github.com/DeepMark}
\BIBentrySTDinterwordspacing

\bibitem{hk_benchmark_2017}
\BIBentryALTinterwordspacing
``The source code and experimental data of benchmarking state -of-the-art deep
  learning software tools,'' 2017, (Accessed: 17/12/ 2017). [Online].
  Available: \url{http://dlbench.comp.hkbu.edu.hk/}
\BIBentrySTDinterwordspacing

\bibitem{lecun_mnist_1998}
\BIBentryALTinterwordspacing
Y.~LeCun, C.~Cortes, and C.~J. Burges, ``The {MNIST} database of handwritten
  digits,'' 1998, (Accessed on: 17/12/2017). [Online]. Available:
  \url{http://yann.lecun.com/exdb/mnist/}
\BIBentrySTDinterwordspacing

\bibitem{zaremba_rnn_2014}
W.~Zaremba, I.~Sutskever, and O.~Vinyals, ``Recurrent neural network
  regularization,'' \emph{CoRR}, vol. abs/1409.2329, 2014.

\bibitem{sermanet_overfeat_2013}
P.~Sermanet, D.~Eigen, X.~Zhang, M.~Mathieu, R.~Fergus, and Y.~LeCun,
  ``{OverFeat}: {Integrated} {Recognition}, {Localization} and {Detection}
  using {Convolutional} {Networks},'' \emph{CoRR}, vol. abs/1312.6229, 2013.

\bibitem{murphy_benchmark_2017}
\BIBentryALTinterwordspacing
J.~Murphy, ``Deep learning benchmarks of {NVIDIA Tesla P100 PCIe}, {Tesla K80},
  and {Tesla M40 GPUs},'' Jan 2017, (Accessed on: 17/12/2017). [Online].
  Available:
  \url{https://www.microway.com/hpc-tech-tips/deep-learning-benchmarks-nvidia-tesla-p100-16gb-pcie-tesla-k80-tesla-m40-gpus/}
\BIBentrySTDinterwordspacing

\bibitem{chollet_dl_lim_2017}
\BIBentryALTinterwordspacing
F.~Chollet, ``The limitations of deep learning,'' 2017, (Accessed on:
  12/12/2017). [Online]. Available:
  \url{https://blog.keras.io/the-limitations-of-deep-learning.html}
\BIBentrySTDinterwordspacing

\bibitem{zenil_causal_reprogramming_2017}
H.~Zenil and {et al.}, ``\BIBforeignlanguage{en}{An {Algorithmic} {Information}
  {Calculus} for {Causal} {Discovery} and {Reprogramming} {Systems}},''
  \emph{\BIBforeignlanguage{en}{bioRxiv}}, p. 185637, 2017.

\bibitem{shwartz-ziv_bb_dnn_2017}
R.~Shwartz-Ziv and N.~Tishby, ``Opening the {Black} {Box} of {Deep} {Neural}
  {Networks} via {Information},'' \emph{CoRR}, vol. abs/1703.00810, Mar. 2017.

\bibitem{nguyen_dl_fool_2015}
A.~M. Nguyen, J.~Yosinski, and J.~Clune, ``Deep neural networks are easily
  fooled: High confidence predictions for unrecognizable images,'' in
  \emph{Proc. CVPR}, 2015, pp. 427--436.

\bibitem{szegedy_ipnn_2014}
\BIBentryALTinterwordspacing
C.~Szegedy, W.~Zaremba, I.~Sutskever, J.~Bruna, D.~Erhan, I.~Goodfellow, and
  R.~Fergus, ``Intriguing properties of neural networks,'' in
  \emph{International Conference on Learning Representations}, 2014. [Online].
  Available: \url{http://arxiv.org/abs/1312.6199}
\BIBentrySTDinterwordspacing

\bibitem{baker_standardizing_data_2013}
N.~A. Baker, J.~D. Klemm, S.~L. Harper, S.~Gaheen, M.~Heiskanen,
  P.~Rocca-Serra, and S.-A. Sansone, ``\BIBforeignlanguage{En}{Standardizing
  data},'' \emph{\BIBforeignlanguage{En}{Nat. Nanotechnol.}}, vol.~8, no.~2,
  p.~73, 2013.

\bibitem{wittig_data_2017}
U.~Wittig, M.~Rey, A.~Weidemann, and W.~Müller, ``Data management and data
  enrichment for systems biology projects,'' \emph{J. Biotechnol.}, vol. 261,
  pp. 229--237, 2017.

\bibitem{mahmud_soa_2012}
M.~Mahmud, M.~M. Rahman, D.~Travalin, P.~Raif, and A.~Hussain, ``Service
  {Oriented} {Architecture} {Based} {Web} {Application} {Model} for
  {Collaborative} {Biomedical} {Signal} {Analysis},'' \emph{Biomed. Tech.
  (Berl).}, vol.~57, pp. 780--783, 2012.

\bibitem{mahmud_webqst_2014}
M.~Mahmud, R.~Pulizzi, E.~Vasilaki, and M.~Giugliano, ``A {Web}-{Based}
  {Framework} for {Semi}-{Online} {Parallel} {Processing} of {Extracellular}
  {Neuronal} {Signals} {Recorded} by {Microelectrode} {Arrays},'' in
  \emph{Proc. MEAMEETING}, 2014, pp. 202--203.

\bibitem{angelov_dl_challenges_2016}
P.~Angelov and A.~Sperduti, ``Challenges in deep learning,'' in \emph{Proc.
  ESANN}, 2016, pp. 489--495.

\bibitem{arulkumaran_deep_rl_2017}
K.~Arulkumaran, M.~P. Deisenroth, M.~Brundage, and A.~A. Bharath, ``Deep
  {Reinforcement} {Learning}: {A} {Brief} {Survey},'' \emph{IEEE Signal Process
  Mag.}, vol.~34, no.~6, pp. 26--38, 2017.

\end{thebibliography}

\end{document}